\documentclass[aps,twocolumn,preprintnumbers,nofootinbib]{revtex4-1}
\usepackage[a4paper, hdivide={1.91cm,,1.165cm}, vdivide={1.83cm,,3.8cm}]{geometry}
\usepackage{capt-of}
\usepackage{ragged2e}
\usepackage{color}
\usepackage[hyperfootnotes=false,colorlinks,citecolor=magenta]{hyperref}
\usepackage{ulem}
\usepackage{youngtab}
\usepackage[english]{babel} 
\usepackage{amsmath}
\usepackage{float}
\usepackage{amssymb}    
\usepackage{braket}
\usepackage{amsfonts,bbm}
\usepackage{dsfont,tensor}
\usepackage{enumerate}
\makeatletter\AtBeginDocument{\let\@elt\relax}\makeatother
\usepackage{graphicx}
\usepackage{caption}
\usepackage{adjustbox}
\usepackage{bbm}
\usepackage{array}
\usepackage{booktabs}
\usepackage{units}
\usepackage{textcomp}
\usepackage{mathtools}
\usepackage{cancel}
\usepackage{slashed}
\usepackage{multirow}
\usepackage{slashed}
\usepackage{comment}
\usepackage{relsize}
\usepackage{tcolorbox}
\usepackage{csquotes}
\usepackage{simpler-wick}
\usepackage{subcaption}
\usepackage{lipsum}
\usepackage[section]{placeins}

\allowdisplaybreaks

\definecolor{myRed}{RGB}{205, 126, 143}
\definecolor{myGreen}{RGB}{138, 185, 155}
\definecolor{myGray}{RGB}{128, 128, 128}
\definecolor{myViolet}{RGB}{164, 140, 177}
\definecolor{amber}{rgb}{1.0, 0.75, 0.0}

\newcommand{\pts}[1]{{.}\hfill(\textit{#1~point}\ifthenelse{\equal{#1}{1}}{}{\textit{s}})}

\newcommand{\Msun}{{\ifmmode{{\rm{M_{\odot}}}}\else{${\rm{M_{\odot}}}$}\fi}}

\newcommand{\beq}{\begin{equation}}
\newcommand{\eeq}{\end{equation}}
\newcommand{\bea}{\begin{eqnarray}}
\newcommand{\ena}{\end{eqnarray}}

\newcommand{\lsim}{\mathrel{\mathop{\kern 0pt \rlap
{\raise.2ex\hbox{$<$}}}
\lower.9ex\hbox{\kern-.190em $\sim$}}}
\newcommand{\gsim}{\mathrel{\mathop{\kern 0pt \rlap
{\raise.2ex\hbox{$>$}}}
\lower.9ex\hbox{\kern-.190em $\sim$}}}

\begin{document}

\title{Dwarf Galaxy Constraints on Interacting Fermionic Dark Matter}
\author{Bihag Dave}
\email{bihag@physics.iitm.ac.in}
\affiliation{Centre for Strings, Gravitation and Cosmology, Department of Physics, Indian Institute of Technology Madras, Chennai 600036, India}

\author{Raghuveer Garani}
\email{garani@iitm.ac.in}
\affiliation{Centre for Strings, Gravitation and Cosmology, Department of Physics, Indian Institute of Technology Madras, Chennai 600036, India}

\begin{abstract}
    Dwarf galaxies in the Local Group offer a way to test dark matter (DM) models against stellar kinematic data. In this work, we study degenerate fermionic DM in two cases: the standard non-interacting Fermi gas, and an interacting degenerate DM fluid described by a phenomenological equation of state motivated by interacting Fermi systems. These interactions modify the compressibility of the DM fluid and, in some regions of parameter space, lead to mechanically unstable branches that must be treated through a Maxwell construction. We solve the corresponding non-relativistic hydrostatic equations consistently and compute the line-of-sight velocity-dispersion profiles using the spherical Jeans equation. We then perform MCMC fits to eight classical Milky Way dwarf spheroidal galaxies. The data favor DM fermion masses in the range $100$--$300\,{\rm eV}$. We find that the interacting and non-interacting equations of state give broadly similar posterior distributions for the fermion mass, central density, and stellar anisotropy. Current data therefore do not strongly prefer an interacting equation of state over the free degenerate Fermi-gas, thereby excluding large deviations from the non-interacting limit.
\end{abstract}
\maketitle

\section{Introduction}
Understanding the microscopic nature of dark matter (DM) remains one of the central open problems in contemporary physics. While the standard cold dark matter (CDM) model has been remarkably successful at describing structure formation on large scales, and is an integral ingredient of the $\Lambda$CDM model of cosmology, the fundamental properties of the DM particle have yet to be pinned down~\cite{ParticleDataGroup:2022pth}. Many DM models that appear CDM-like at large scales produce qualitatively different signatures at smaller scales, for instance in the internal structure of dark matter halos, allowing us to probe their mass, spin and self-couplings~\cite{Tulin:2017ara, Arguelles:2023nlh, Cirelli:2024ssz, Eberhardt:2025caq}. Dwarf galaxies, in particular, satellites of the Milky Way and Andromeda (M31), provide a unique laboratory to probe dark matter properties. These systems are strongly dark matter dominated, and characterized by relatively low baryonic content~\cite{Walker:2009, Strigari:2018utn, Battaglia:2022Nat}. 

The primary observational parameter for these systems is the line-of-sight (LOS) velocity distribution of their member stars. High-resolution spectroscopic surveys have enabled precise measurements of LOS velocities for hundreds to thousands of stars in classical dwarf spheroidals~\cite{Walker:2007ju, Simon:2007dq}, and tens of stars in ultra-faint systems~\cite{Simon_2019}. From these data, both global velocity dispersions and, in some cases, radially resolved dispersion profiles can be constructed.  A robust outcome of such studies is that dwarf galaxies exhibit velocity dispersions of  $\mathcal{O}(10)\,\mathrm{km\,s^{-1}}$, with dynamical mass-to-light ratios that can exceed $10^2$--$10^3$~\cite{Simon_2019, Battaglia:2022Nat,Pascale:2025zga}. These observations provide compelling evidence that dwarf galaxies are dominated by dark matter at all radii accessible to stellar kinematic probes.

The interpretation of LOS kinematics requires connecting projected velocity moments to the underlying phase-space distribution of DM and stars, essentially probing the gravitational potential. While ``global" quantities, such as the velocity dispersion within the half-light radius, provide largely model-independent estimates of the enclosed mass~\cite{Wolf:2009tu}, radially resolved kinematic profiles provide additional information about the structure of the gravitational potential and the orbital distribution of stars. However, the extraction of this information is limited by well-known degeneracies, most notably between the mass profile and the baryonic velocity anisotropy~\cite{BinneyMamon:1982}.

Even with current limitations, stellar kinematics of dwarf galaxies have enabled probes of various dark matter models that have non-trivial implications at galactic scale in recent years~\cite{Correa:2020qam, Domcke:2014kla, Boyarsky:2008ju, Diez-Tejedor:2014naa, Goldstein:2022pxu, Korshynska:2025nia, Wardana:2026qag, Savchenko:2019qnn, DiPaolo:2017geq, Alvey:2020xsk, Ando:2025qtz}. In this work, we focus on fermionic DM with non-negligible self-interactions. Fermionic candidates are an interesting possibility as degeneracy pressure can stabilize gravitational collapse and generate extended cores. In its simplest realization, this corresponds to a non-relativistic degenerate Fermi gas, characterized by a polytropic equation of state $P \propto \rho^{5/3}$~\cite{Boyarsky:2008ju, Domcke:2014kla, Savchenko:2019qnn, DiPaolo:2017geq, Bar:2021jff, Chavanis:2021jds}.\footnote{Dwarf spheroidal phase-space constraints exclude arbitrarily light fermionic dark matter, but the model-independent bounds are relatively weak, typically $m_{\rm DM}\gtrsim 0.1\text{--}0.2\,{\rm keV}$ \cite{Domcke:2014kla,DiPaolo:2017geq,Alvey:2020xsk,Savchenko:2019qnn}.  Stronger, keV-scale lower bound, e.g. $m\gtrsim 1.7\,{\rm keV}$, apply only for specific production mechanisms such as non-resonant sterile-neutrino dark matter~\cite{Boyarsky:2008ju, Bezrukov:2025ttd}.} 
While this minimal scenario captures the essential role of fermi pressure, it does not allow for the possibility of additional interactions in the dark sector. In this work, we go beyond the non-interacting degenerate fermi gas scenario by considering a class of phenomenological equations of state (EoS) that take into account attractive self-interactions. Such interactions can arise, for instance, from Yukawa-like forces in the dark sector~\cite{Walecka:1974qa,Gresham:2018rqo,Garani:2022quc}.  The central question of this work is therefore to test whether current data are sensitive to qualitatively new equation-of-state (EoS) features beyond the free degenerate Fermi gas. In particular, we ask whether attractive interactions, which soften the equation of state and can give rise to phase coexistence at strong coupling, leave observable imprints on halo structure. 

This paper is organized as follows. In Sec.~\ref{sec:dwarfs}, we summarize the available kinematic data for Local Group dwarf galaxies. In Sec.~\ref{sec:eos}, we introduce the phenomenological equation of state considered in this work. In Sec.~\ref{sec:hydro}, we describe the construction of hydrostatic equilibrium configurations and their properties.  In Sec.~\ref{sec:results}, we present MCMC analysis and results. We conclude in Sec.~\ref{sec:conclusions} with a discussion on the implications for DM micro-physics.


\section{Dwarf spheroidal galaxies}\label{sec:dwarfs}

Local Group dwarf galaxies provide two qualitatively distinct kinds of stellar-kinematic observables. The first is the radially resolved line-of-sight (LOS) velocity-dispersion profile, $\sigma_{\rm los}(R)$, which probes the radial structure of the gravitational potential. The second is the global line-of-sight velocity dispersion, which yields a robust estimate of the enclosed mass near the half-light radius and is therefore especially valuable in sparsely sampled systems. 

The local LOS velocity dispersion is defined as a function of projected radius $R$ on the plane of the sky,
\begin{equation}
\sigma_{\rm los}^2(R) = \langle (v_{\rm los} - \bar{v})^2 \rangle_R,
\end{equation}
where the average is taken over stars within an annulus at projected radius $R$. Observationally, stars are binned radially, membership is assigned probabilistically, and measurement uncertainties are incorporated into likelihood-based estimators~\cite{Walker:2005nt, Walker:2007ju}. For classical dwarf spheroidals, sufficiently large spectroscopic samples allow the construction of smooth radial profiles, whereas for ultra-faint dwarfs such profiles are often difficult to construct.

In this section we summarize these observables and clarify which data are sufficiently informative for the analysis carried out in this paper.

\subsection{Jeans Analysis}\label{subsec:jeans}

The observable $\sigma_{\rm los}(R)$ is related to dynamical structure through the spherical Jeans equation~\cite{BinneyTremaine:2008, Walker:2007ju},
\begin{equation}\label{eq:jeans}
\frac{d}{dr}\left( \rho_*(r)\,\sigma_r^2(r) \right) + \frac{2\beta(r)}{r}\,\rho_*(r)\,\sigma_r^2(r)
= -\rho_*(r)\frac{G M(r)}{r^2},
\end{equation}
where $\rho_*(r)$ is the 3D stellar density, $\sigma_r(r)$ the stellar radial velocity dispersion, and $\beta(r) = 1 - \sigma_\theta^2/\sigma_r^2$ the stellar velocity anisotropy.\footnote{Physically, $\beta\rightarrow \infty$ indicates that the orbits are near circular, isotropic if $\beta \rightarrow 0$. Orbits are radial 
if $\beta=1$. Note that negative values of $\beta$ signifies large tangential component.}
$M(r)$ is the total mass enclosed within radius $r$ which, for dark matter dominated dwarf spheroidals, can be estimated by the mass profile of dark matter inside the halo (see sec.~\ref{sec:hydro}). 
The Jeans equations has many classes solutions. 
For the physically relevant case of constant $\beta$ the solution reads~\cite{BinneyMamon:1982,Mamon:2004xk}
\begin{equation}\label{eq:rhostar}
    \rho_*(r)\,\sigma_r^2(r) = G r^{-2 \beta}\int_r^\infty x^{2\beta -2} \rho_\ast(x) M(x) d x~,
\end{equation}

Projection along the line of sight yields (for constant $\beta$) the following 1D velocity dispersion~\cite{BinneyMamon:1982}
\begin{equation}\label{eq:vel_disp}
\sigma_{\rm los}^2(R) = \frac{2}{\Sigma_*(R)} \int_R^\infty 
\left(1 - \beta\frac{R^2}{r^2}\right)
\frac{\rho_*(r)\,\sigma_r^2(r)\,r}{\sqrt{r^2 - R^2}}\,dr~,
\end{equation}
with $\Sigma_*(R)$ the projected stellar density, often chosen to be the Plummer profile~\cite{Belokurov:2007}. This reads
\begin{equation}
    \Sigma_*(R) = \frac{L}{\pi r_{\text{half}}^2} \frac{1}{(1+ R^2/r_{\text{half}}^2)^2}~, 
\end{equation}
where $L$ is the total luminosity and $r_{\text{half}}$ the 2D projected half-light radius.
Given the projected stellar density, in this case the Plummer profile, the 3D stellar density profile $\rho_*(r)$ is given by~\cite{Walker:2009}, 

\begin{equation}
    \rho_*(r) = -\frac{1}{\pi}\int^\infty_r \frac{\mathrm{d}\Sigma_\star}{\mathrm{d}r} \frac{\mathrm{d}R}{\sqrt{R^2 - r^2}} = \frac{3L}{4\pi r_{\text{half}}^3}\frac{1}{\left[1 + (r/r_{\text{half}})^2\right]^{5/2}}\ .
\end{equation}
We take the values of $L$ and $r_{\text{half}}$ from Refs.~\cite{Battaglia:2022Nat, Walker:2009}. Empirically, the LOS dispersion profiles of most dwarf galaxies are approximately flat with radius, with occasional mild gradients and statistically significant fluctuations. These features encode information about the mass profile $M(r)$, the anisotropy $\beta(r)$, and possible departures from equilibrium such as tidal effects or mergers. However, a central limitation is the mass--anisotropy degeneracy: different combinations of $M(r)$ and $\beta(r)$ can produce indistinguishable $\sigma_{\rm los}(R)$, making the inference of inner density slopes and DM properties inherently model-dependent.

\subsubsection{Global LOS}
In contrast, the global LOS velocity dispersion is defined as the variance of all member star velocities,
\begin{equation}
\sigma_{\rm los}^2 = \langle (v_{\rm los} - \bar{v})^2 \rangle,
\end{equation}
and is typically the most robust observable available. A key result in dwarf galaxy dynamics is the existence of a simple estimator for the dynamical mass within the half-light radius~\cite{Battaglia:2013wqa},
\begin{equation}
M(r_{1/2}) \simeq \frac{4\,\sigma_{\rm los}^2\,r_{\text{half}}}{G}.
\end{equation}

Note that $r_{1/2}$ is 3D half-light radius while $r_{\rm half}$ is 2D projected half-light radius. This relation depends only weakly on the anisotropy and holds across a broad class of dynamical models, making it particularly useful for systems with limited data. Observationally, dwarf galaxies exhibit LOS dispersions in the range $\sim 2$--$12~\mathrm{km\,s^{-1}}$~\cite{dwarflocalgroup:2012}, corresponding to mass-to-light ratios from $\sim 10$ up to $\sim 10^3$. This implies that these systems are dark-matter dominated even within their half-light radii making such galaxies an ideal probe of the dark matter distribution within them.

Global LOS measurements are robust to baryonic modelling assumptions and applicable even to sparsely populated ultra-faint systems, however, they sacrifice spatial information. In particular, they cannot constrain the radial variation of the mass profile, distinguish between cored and cuspy halos, or reveal kinematic substructure. From a theoretical perspective, global LOS dispersion provides a reliable normalization of the gravitational potential at the half-light scale, while local LOS profiles offer a window into the radial structure of dark matter halos, albeit being model-dependent.

\subsection{Data Sets used in this work}

We use LOS dispersion data for the classical MilkyWay dwarf spheroidal galaxies Draco, Sculptor, Fornax, Sextans, Carina, and Ursa Minor, presented in refs.~\cite{Walker:2005nt, Walker:2007ju, Walker:2009}. Each of these dwarfs spheroidals have been studied with large spectroscopic samples ranging from $\sim 200$ up to $\sim 2500$ member stars. These datasets enable the construction of radially resolved velocity dispersion profiles extending out to several half-light radii~\cite{Walker:2009}. Observationally, these systems exhibit LOS dispersion profiles that are approximately flat as a function of projected radius, with values typically in the range $\sigma_{\rm los} \sim 6$--$12~\mathrm{km\,s^{-1}}$~\cite{Battaglia:2022Nat}. Deviations from strict flatness are present in the form of mild radial gradients and fluctuations, which are not fully attributable to measurement uncertainties and may reflect non-trivial dynamics or the presence of multiple stellar components.

In addition to dispersion profiles, searches for velocity gradients in Milky Way dwarfs have studied to identify ordered motions or tidal effects. Most systems show little or no significant rotation indicating that random motions dominate~\cite{Battaglia:2022Nat}. However, exceptions exist: for example, Antlia II and Tucana III show evidence for velocity gradients that may be associated with tidal disruption, while systems such as Draco, Fornax, Ursa Minor and Carina exhibit weak gradients at low significance~\cite{Martinez_Garcia_2022}. Importantly, the detection of such gradients depends sensitively on radial coverage, as the amplitude of velocity gradients typically increases toward larger radii~\cite{Battaglia:2022Nat}.

\subsection{Limitations of current M31 data}
For satellites of M31, the situation is more heterogeneous. Due to their larger distances, spectroscopic samples are typically smaller, often limited to $\sim 20$--$200$ member stars. As a result, radially resolved $\sigma_{\rm los}(R)$ profiles are available only for a subset of relatively bright systems such as And II~\cite{AndromedaII:2012,AndromedaII:2017} and And XIX~\cite{Collins_AndXIX:2020}
Even in these cases, the radial coverage is often incomplete, and uncertainties in the outer regions are substantial. Nevertheless, the available data suggest that M31 dwarfs occupy similar regions of parameter space as Milky Way dwarfs in terms of their velocity dispersions and structural properties, with typical values $\sigma_{\rm los} \sim 6$--$12~\mathrm{km\,s^{-1}}$ and broadly flat or weakly varying dispersion profiles~\cite{Battaglia:2022Nat}.

Some M31 dwarfs exhibit distinctive kinematic features. Andromeda II is a notable case, displaying significant rotation, including evidence for prolate rotation (rotation around the major axis), which appears rare for systems of this type~\cite{AndromedaII:2012}. This rotation is accompanied by variations in kinematic properties with stellar population, suggesting a complex formation htistory possibly involving mergers. Similarly, And XIX shows an unusually large spatial extent combined with relatively low velocity dispersion, which may indicate tidal effects or an atypical dark matter distribution~\cite{Collins_AndXIX:2020}. However, for many other M31 dwarfs, the available data are insufficient to robustly detect such features.

Among the presently available M31 systems, And XIX is of particular interest because a radial velocity-dispersion profile has been reported~\cite{Collins_AndXIX:2020}. However, the profile contains only a small number of radial bins, and this substantially limits the constraining power of a multi-parameter fit. In such cases, a full Jeans analysis with several free halo and anisotropy parameters risks becoming over-parameterised. At the present, we therefore, do not utilize this LOS information.

A key limitation affecting both Milky Way and M31 systems is the difficulty of obtaining accurate measurements in the outer regions of galaxies, where membership determination becomes increasingly uncertain and contamination from foreground stars becomes significant. While techniques based on metallicity, surface gravity indicators, and, more recently, Gaia proper motions have improved membership selection, residual contamination can still bias the inferred dispersion profiles, particularly at large radii~\cite{Battaglia:2022Nat,Martin_pandas:2026}. Nonetheless, studies indicate that such contamination has only a minor impact on global dispersion values, though it may affect the detailed shape of $\sigma_{\rm los}(R)$.

In summary, Milky Way and M31 dwarf galaxies differ sharply in the quality of available stellar-kinematic data. For Milky Way satellites, high-quality, radially resolved LOS kinematics provide detailed constraints on internal dynamics, revealing flat dispersion profiles, multiple stellar components, and occasional velocity gradients. For M31 dwarfs, the data are more limited, allowing only partial reconstruction of $\sigma_{\rm los}(R)$ in the best cases and restricting most systems to global measurements. Despite these differences, the available evidence suggests that both populations share broadly similar dynamical properties, consistent with a common underlying DM dominated structure.\\

\section{Phenomenological equation of state for Dark Matter}\label{sec:eos}

\noindent\underline{\it Degenerate Fermi gas:}  The EoS for a non-relativistic, fully degenerate Fermi gas provides a natural baseline for fermionic dark matter supported by Fermi pressure. This reads 
\begin{eqnarray}\label{eq:eos_fermi}
    P(\rho) = K \rho^{5/3} ~, 
\end{eqnarray}
with $K = \frac{h^2}{5m_\chi^{8/3}}\left(\frac{3}{8\pi}\right)^{2/3}$. Its structure is well known from applications ranging from white dwarfs to neutron stars~\cite{Glendenning_2000}, and follows directly from filling momentum states up to the Fermi surface in the absence of interactions. 
The polytropic index $n=3/2$ (or equivalently $\gamma = 5/3$ used in the above equation) implies a relatively stiff equation of state, leading to extended cores whose size is determined primarily by the fermion mass $m_\chi$ as well as the central density $\rho_c$. 
This framework has been widely explored as a minimal realization of fermionic dark matter halos, where phase-space considerations (and LOS velocity dispersions~\cite{Domcke:2014kla}) impose lower bounds on $m_\chi$ through the Pauli exclusion principle or equivalently the Tremaine-Gunn bound (see e.g. \cite{Tremaine:1979we, Ruffini:1983zz, Boyarsky:2008ju}). 

While the degenerate Fermi gas captures the essential role of degeneracy pressure, it neglects any interactions in the dark sector. From a bottom-up perspective, this is a simplifying assumption. It is therefore useful to consider controlled modifications of Eq.~\eqref{eq:eos_fermi} that incorporate interactions in a phenomenological manner.

\medskip

\noindent\underline{\it Interacting Fermi gas:} 
Motivated by the Thomas--Fermi approach to interacting fermionic systems, we introduce a phenomenological equation of state that captures the impact of attractive interactions, for instance, mediated by Yukawa\footnote{See appendix~\ref{app:UV_model}, where microphysical origins of such EoS are derived within a simple model. The phenomenological equation of state in Eq.~\eqref{eq:eos_ph} can be interpreted as arising from a coarse-grained phase-space distribution of fermions in a density-dependent mean-field potential, analogous to the Thomas–Fermi approximation in many-body systems.} interactions~\cite{Walecka:1974qa, Gresham:2018rqo, Garani:2022quc}, given by, 
\begin{eqnarray}\label{eq:eos_ph}
    P(\rho) = K \rho^{5/3} - \alpha\frac{\rho^2}{\rho_s^2 + \rho^2}~. 
\end{eqnarray}

Here $\rho_s$ defines a characteristic density scale at which finite-density effects become relevant, while $\alpha$ parametrizes the effective strength of the attractive interaction. The functional form of the interaction term is chosen to interpolate smoothly between two regimes: at low densities ($\rho/\rho_s \ll 1$), the correction scales approximately as $-(\rho/\rho_s)^2$, while at high densities ($\rho/\rho_s \gg 1$) it saturates, leading to a $-\alpha$ correction term. In the large density regime, the EoS resembles that of a simple Bag model~\cite{Alford:2004pf} in the non-relativistic limit. 

This saturating form prevents the unphysical runaway collapse of DM and naturally gives rise to a self-bound configuration with a well-defined equilibrium density, analogous to the saturation of nuclear matter. In the present context it manifests as a finite, stable dark matter core whose size and central density are governed by the interplay between the degeneracy pressure and the self-interaction strength. This structure is reminiscent of screening effects in many-body systems, where interactions become less efficient due to finite-density effects~\cite{Walecka:1974qa}. While remaining agnostic about the precise ultraviolet completion of the underlying DM theory, the phenomenological EoS considered here provides a useful description that can be tested conveniently with available data.

\begin{figure*}[htb!]
    \centering
    \includegraphics[width=0.94\textwidth]{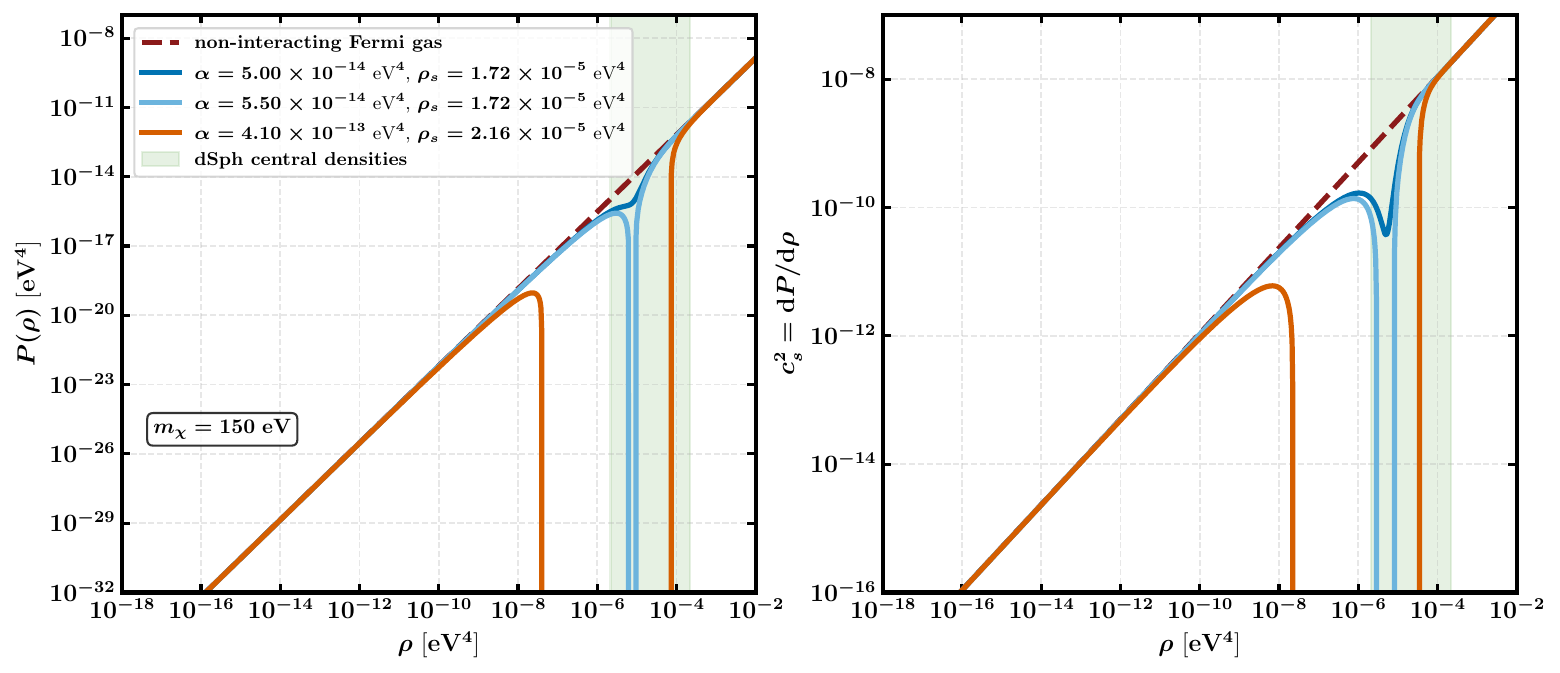}
    \caption{\justifying \textit{Left:} Illustrative examples of various equations of state shown for a fixed $m_\chi = 150\ \text{eV}$ and for various values of $\alpha$ and $\rho_s$. \textit{Right:} For the same parameters, sound speed $c_s^2 = \mathrm{d}P/\mathrm{d}\rho$ is also shown. Typical values of the central densities for the degenerate Fermi gas model for dwarf spheroidal galaxies are also shown by the shaded green region~\cite{Walker:2009,Domcke:2014kla}.}
    \label{fig:eos_comparison}
\end{figure*}

Examples for such equations of state are shown in fig. (\ref{fig:eos_comparison}). For phenomenological applications to dwarf galaxies, values such as $\rho_s \sim 1~\mathrm{GeV/cm^3}$ are motivated by typical central densities inferred from kinematic data, while fermion masses in the range $m_\chi \sim \mathcal{O}(10^2~\mathrm{eV})$ provide core sizes ($\sim {\rm kpc}$) compatible with observations. Within this parameter space, the interplay between degeneracy pressure and attractive interactions leads to qualitatively new features in the equation of state. In particular, there exist parametric regions, light blue curve in fig.~(\ref{fig:eos_comparison}), where the pressure becomes negative and, more importantly, where the compressibility turns negative, $dP/d\rho < 0$. The latter signals a mechanical instability, analogous to the liquid--gas instability in ordinary matter. In such instances, it is appropriate to interpret Eq.~\eqref{eq:eos_ph} as describing a metastable branch, and to replace the unstable segment with a thermodynamically consistent construction. Following standard practice, we implement a Maxwell construction, which enforces phase coexistence between two densities at equal pressure and chemical potential. This is detailed in appendix~\ref{app:maxwell_construction}.  This procedure effectively replaces the region of negative compressibility with a constant-pressure plateau, corresponding to phase co-existence in the dark sector (see, for instance,~\cite{Shapiro:1983du, Gresham:2018rqo} for related discussions in astrophysical contexts). While the precise microscopic origin of such a transition is model-dependent, its macroscopic imprint on halo structure can be studied.

\section{Hydrostatic configurations}\label{sec:hydro}

In the non-relativistic limit, given an equation of state, hydrostatic configurations are obtained by solving the following equations for pressure and energy density 
    \begin{eqnarray}
        \frac{\mathrm{d}\rho(r)}{\mathrm{d}r} &=& -\frac{1}{c_s^2}\frac{GM(r)}{r^2}\rho(r)\ , \label{eq:equilibrium}\\ 
        \frac{\mathrm{d}M(r)}{\mathrm{d}r} &=& 4\pi r^2\rho(r)\ .\label{eq:continuity}
    \end{eqnarray}
The sound speed is defined as $c_s^2 \equiv \mathrm{d}P/\mathrm{d}\rho$. 
The equations are adimensionalized by rewriting using the following variables: $\rho = \rho_c \theta(\xi)$, $M  = M_c y(\xi)$, where $r = \alpha \xi$, $\alpha \equiv \left(c_s^2(\rho_c)/(4\pi G \rho_c)\right)^{1/2}$ and $M_c \equiv 4\pi \rho_c \alpha^3$. 
Here, we have used $\rho_c$ to denote the central density of the equilibrium configuration.
We also use the definition of sound speed to define $K(\theta) \equiv c_s^2(\rho)/c_s^2(\rho_c)$. Note that $K(1) = 1$.
Using the boundary conditions, $\theta(0) = 1$ (which corresponds to $\rho(0) = \rho_c$) and $y(0) = 0$,  the following equations are solved numerically
    \begin{eqnarray}\label{eq:int_LE1}
        \frac{\mathrm{d}\theta(\xi)}{\mathrm{d}\xi} &=& -\frac{y(\xi)\theta(\xi)}{\xi^2 K(\theta)}\ , \\ \label{eq:int_LE2}
        \frac{\mathrm{d}y(\xi)}{\mathrm{d}\xi} &=& \xi^2 \theta(\xi)\ .
    \end{eqnarray}

In the absence of interactions, the simpler polytropic equation of state (with $\gamma = 5/3$) allows us to write sound speed as simply $c_s^2 = \frac{5}{3} K\rho^{2/3}$. This futher allows us to use the general redefinition $\rho \equiv \rho_c\theta^n$ and the two equations reduce to the Lane-Emden equation:

\begin{equation}\label{eq:LE}
    \frac{1}{\xi^2}\frac{\mathrm{d}}{\mathrm{d}\xi}\left(\xi^2\frac{\mathrm{d}\theta}{\mathrm{d}\xi}\right) = -\theta^n  
\end{equation}
where using $\gamma = 1 + 1/n$ implies the polytropic index $n = 3/2$, which is our baseline model. One can solve this system numerically using the same initial conditions $\theta(0) = 1$ and $\theta'(0) = 0$. For the non-interacting case, the Lane-Emden equation, Eq.~\eqref{eq:LE}, can be solved only once, and then solution can be scaled by appropriate factors of $m_\chi$ and $\rho_c$ to explore different solutions corresponding to different fermion masses and central densities. However, for the interacting scenario, Eqs.~\eqref{eq:int_LE1} and~\eqref{eq:int_LE2}, are solved for each choice of $(m,\alpha,\rho_s,\rho_c)$ since the solution depends on the sound speed $c_s^2$ experienced at densities $\rho_c$ and below. 

\subsection{Solutions to Lane-Emden equation}\label{sec:solutions}
\begin{figure*}[htb!]
    \includegraphics[width = \textwidth]{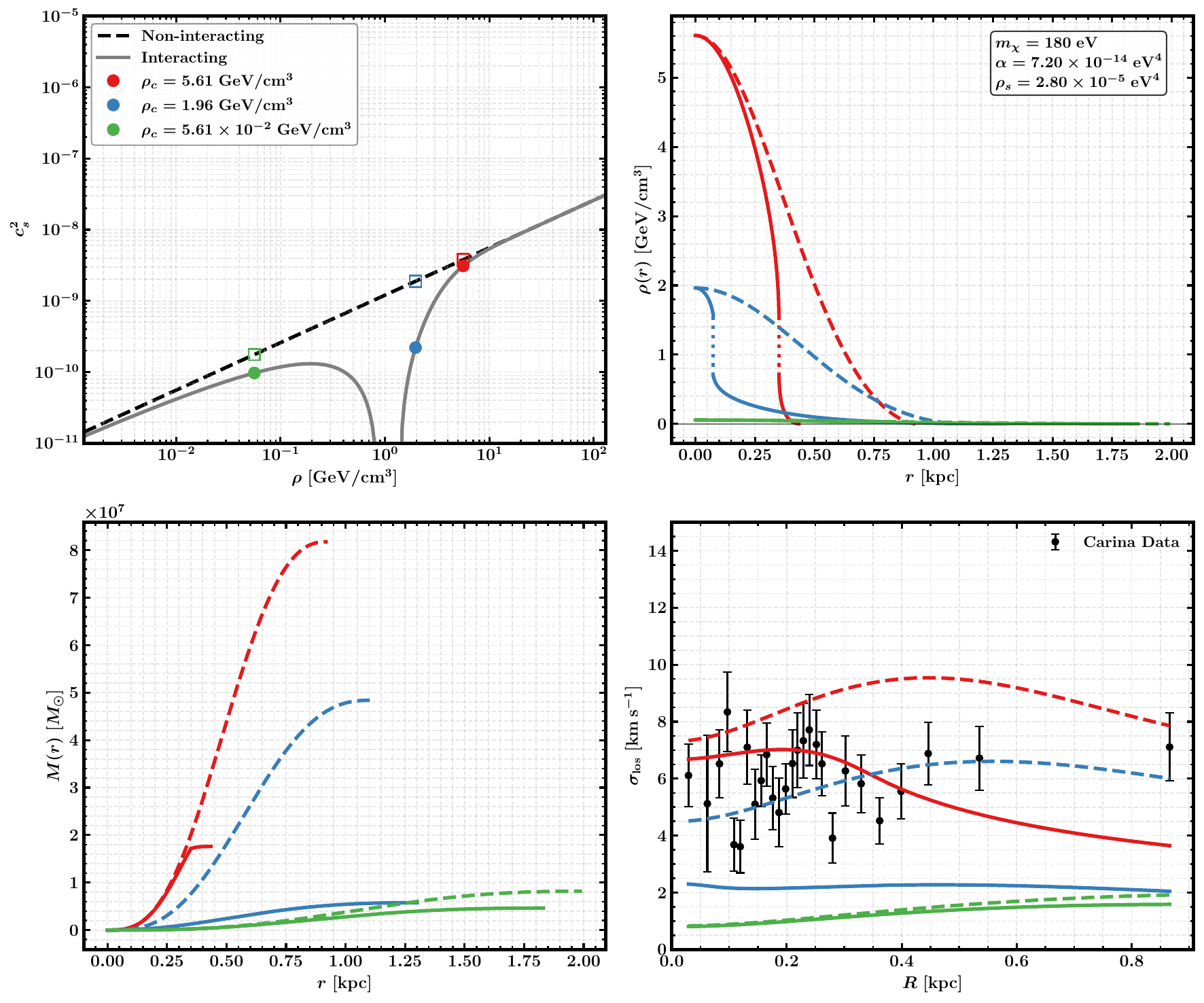}
\caption{\justifying Effect of the interacting EoS on hydrostatic halo structure and projected stellar kinematics for fixed fermion mass, $m_\chi = 180\,{\rm eV}$. {\it Top left:} sound speed squared, $c_s^2=dP/d\rho$, for the non-interacting degenerate Fermi gas (dashed) and for the interacting EoS (solid). The colored markers indicate the central densities used to solve hydrostatic equations. {\it Top right:} resulting density profiles $\rho(r)$; solid curves denote solutions obtained for interacting EoS, while dashed curves show the corresponding non-interacting profiles with the same central densities. The dotted segments represent the phase-transition. {\it Bottom left:} enclosed mass profiles $M(r)$. {\it Bottom right:} predicted line-of-sight velocity-dispersion profiles, obtained from Eq.~\eqref{eq:vel_disp}, compared with the observed stellar kinematic data of the Milky Way dwarf Carina~\cite{Walker:2009}. The interacting EoS suppresses the compressibility over a finite density interval, leading to steeper density profiles and smaller enclosed masses than the non-interacting case for the same central density. Where the solution encounters the mechanically unstable region, the Maxwell construction produces a density discontinuity corresponding to a transition between the two stable branches of the DM fluid.}
    \label{fig:model_comparison}
\end{figure*}
We now illustrate hydrostatic solutions to the phenomenological EoS and discuss how dwarf-galaxy DM halos and, in turn, their LOS velocity dispersion can be affected. As discussed in Sec.~\ref{sec:eos}, the interacting EoS in Eq.~\eqref{eq:eos_ph} can develop non-trivial structure once the attractive interaction becomes sufficiently important. In particular, for large enough $\alpha$, the pressure is reduced over a finite range of densities controlled by the scale $\rho_s$, as shown in fig.~(\ref{fig:eos_comparison}). Since the pressure is no longer described by a single power law over the full density range, the central density $\rho_c$ becomes an important physical parameter: it determines which part of the equation of state is sampled by the hydrostatic solution.

\subsubsection{Fixed DM mass}

We first fix the DM particle mass and study how the halo profile changes as the central density is varied. In fig.~(\ref{fig:model_comparison}), we show representative solutions for
\[
m_\chi = 180\,{\rm eV},  \,\,\,\,\, \alpha = 7.20 \times 10^{-14}\,{\rm eV}^4, \,\,\,\,\, \rho_s = 2.80 \times 10^{-5}\,{\rm eV}^4 .
\]
Throughout the figure, dashed curves denote the non-interacting degenerate Fermi-gas EoS, while solid curves denote the interacting EoS. The different colors correspond to different choices of central density:
\[
\rho_c = 5.61\,{\rm GeV\,cm^{-3}} \quad {\rm(red)}, 
\qquad
\rho_c = 1.96\,{\rm GeV\,cm^{-3}} \quad {\rm(blue)}, 
\]
\[\rho_c = 5.61 \times 10^{-2}\,{\rm GeV\,cm^{-3}} \quad {\rm(green)} .
\]

The top-left panel of fig.~\ref{fig:model_comparison} shows the sound speed squared,
$c_s^2 = dP/d\rho$, for the non-interacting and interacting EoS. The colored markers indicate the central densities used as boundary conditions when solving Eqs.~\eqref{eq:int_LE1} and~\eqref{eq:int_LE2}. This panel is instructive as it shows directly which part of the EoS each hydrostatic solution probes. For the phenomenological EoS in Eq.~\eqref{eq:eos_ph}, the interaction contribution to the pressure is
\[
-\alpha \frac{(\rho/\rho_s)^2}{1+(\rho/\rho_s)^2}.
\]
This contribution is negative and varies most rapidly near $\rho \sim \rho_s$. Consequently, its derivative with respect to $\rho$ suppresses $c_s^2$ relative to the free Fermi-gas value over a finite interval in density. In the limits $\rho \ll \rho_s$ and $\rho \gg \rho_s$, the correction to $c_s^2$ becomes small, and the interacting equation of state approaches the non-interacting behavior.

This suppression of $c_s^2$ has a direct impact on the DM profile. From the hydrostatic equilibrium equation, Eq.~\eqref{eq:equilibrium}, a smaller sound speed at fixed density implies a steeper radial density gradient. Therefore, whenever the solution passes through the softened part of the EoS, the interacting density profile falls more rapidly than the corresponding non-interacting one.

This trend is shown in the top-right panel of fig.~\ref{fig:model_comparison}. For the red and blue curves, the chosen central densities are high enough that the solutions sample the region where the interacting EoS differs significantly from the free Fermi-gas case. The corresponding solid curves therefore show more compact halos, with steeper density profiles and smaller cores than the dashed curves with the same $\rho_c$. By contrast, the green curve starts at a much lower central density, $\rho_c = 5.61 \times 10^{-2}\,{\rm GeV\,cm^{-3}}$, and probes mainly the low-density side of the interacting EoS. In this regime the interaction term behaves approximately as $ -\alpha \rho^2/\rho_s^2$, 
and the profile remains closer to the non-interacting solution. If the central density were taken much larger than $\rho_s$, the inner part of the halo would again resemble the non-interacting case, since the interaction term would be nearly constant and its derivative would be small. The main effect of the interaction would then appear only at larger radii, after the density has fallen into the transition region.

The same trend is visible in the enclosed mass profiles shown in the bottom-left panel of fig.~(\ref{fig:model_comparison}). At fixed central density, the interacting solutions generally enclose less mass within a given radius whenever they pass through the region where the equation of state is softened. This is simply the integrated consequence of the steeper density fall-off induced by the attractive interaction.

For some choices of $(\alpha,\rho_s)$, the sound speed squared becomes negative over a finite density interval. In this case, a continuous hydrostatic solution through that region is not physically allowed, since $c_s^2<0$ signals instability. Equivalently, the right-hand side of Eq.~\eqref{eq:equilibrium} becomes singular at the densities where $c_s^2$ vanishes. As described in appendix~\ref{app:maxwell_construction}, we then replace the unstable portion of the equation of state by the Maxwell-constructed constant-pressure segment. In the density profile, this appears as a discontinuous jump between the two stable branches of the DM fluid.

Numerically, for each choice of $(m_\chi,\alpha,\rho_s,\rho_c)$, we first determine whether the EoS is monotonic. If it is, the hydrostatic equations are integrated outward from the centre in the usual way. If a mechanically unstable segment is present, we first construct the Maxwell-corrected EoS and then integrate the hydrostatic equations. In practice, once the solution reaches the upper coexistence density, the density jumps at fixed pressure to the lower coexistence branch, after which the integration continues on the low-density side. In fig.~(\ref{fig:model_comparison}), the dotted segments indicate this phase-transition region. The green solution lies on the low-density side of the dip in $c_s^2$ and therefore does not encounter the unstable branch. The blue and red solutions begin on the high-density side and pass through the mechanically unstable interval, so that the Maxwell construction is required.

\begin{figure}[htb!]
    \centering
    \includegraphics[width=\linewidth]{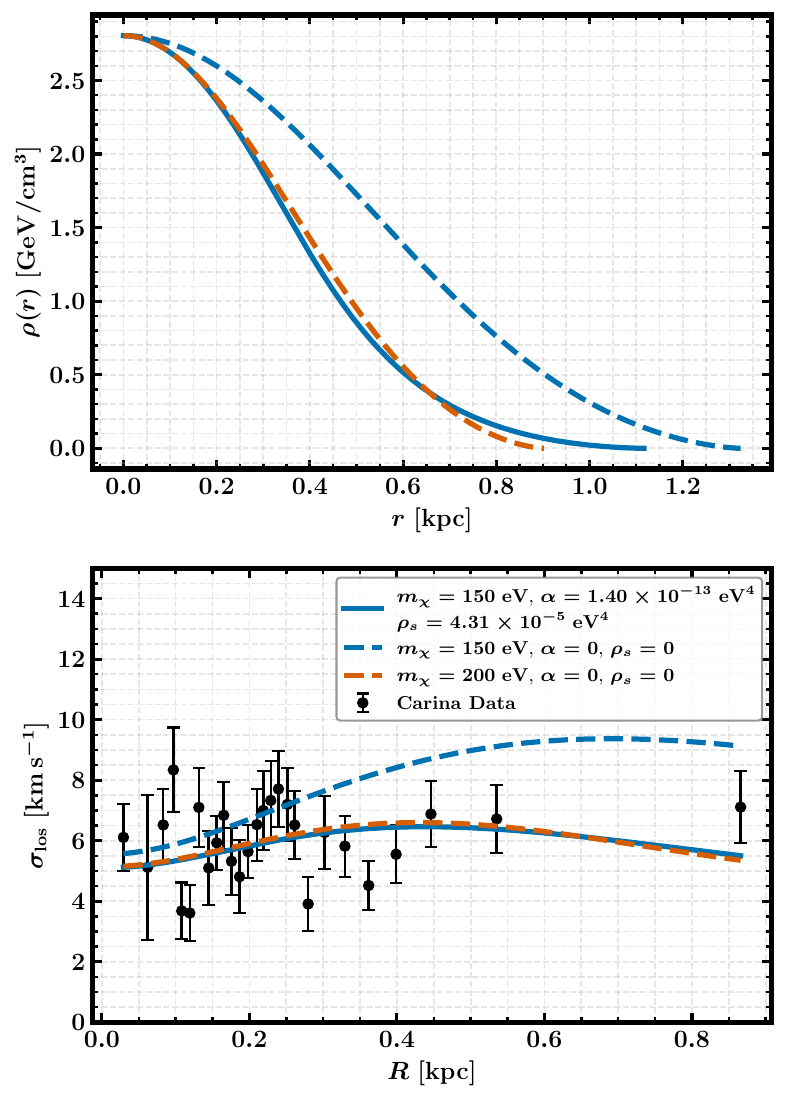}
      \caption{\justifying Example of interplay between fermion mass and attractive self-interactions is illustrated. {\it Top}: DM density profiles $\rho(r)$ for interacting EoS, for DM mass $=150\ \text{eV}$ (solid curve), and DM mass $=200\ \text{eV}$ (dashed curve) without interactions. 
      {\it Bottom}: Corresponding line-of-sight velocity-dispersion profiles $\sigma_{\rm los}(R)$ compared with the observed stellar kinematic data of the Carina dwarf spheroidal. 
      A smaller fermion mass tends to increase the core size in the non-interacting fermi gas EoS, however, attractive interaction steepens the profile. These two effects can partially compensate each other, allowing mildly lighter fermion masses to remain compatible with the observed stellar kinematics.}
    \label{fig:mass_degen}
\end{figure}

\begin{figure*}[htb!]
    \includegraphics[width = \textwidth]{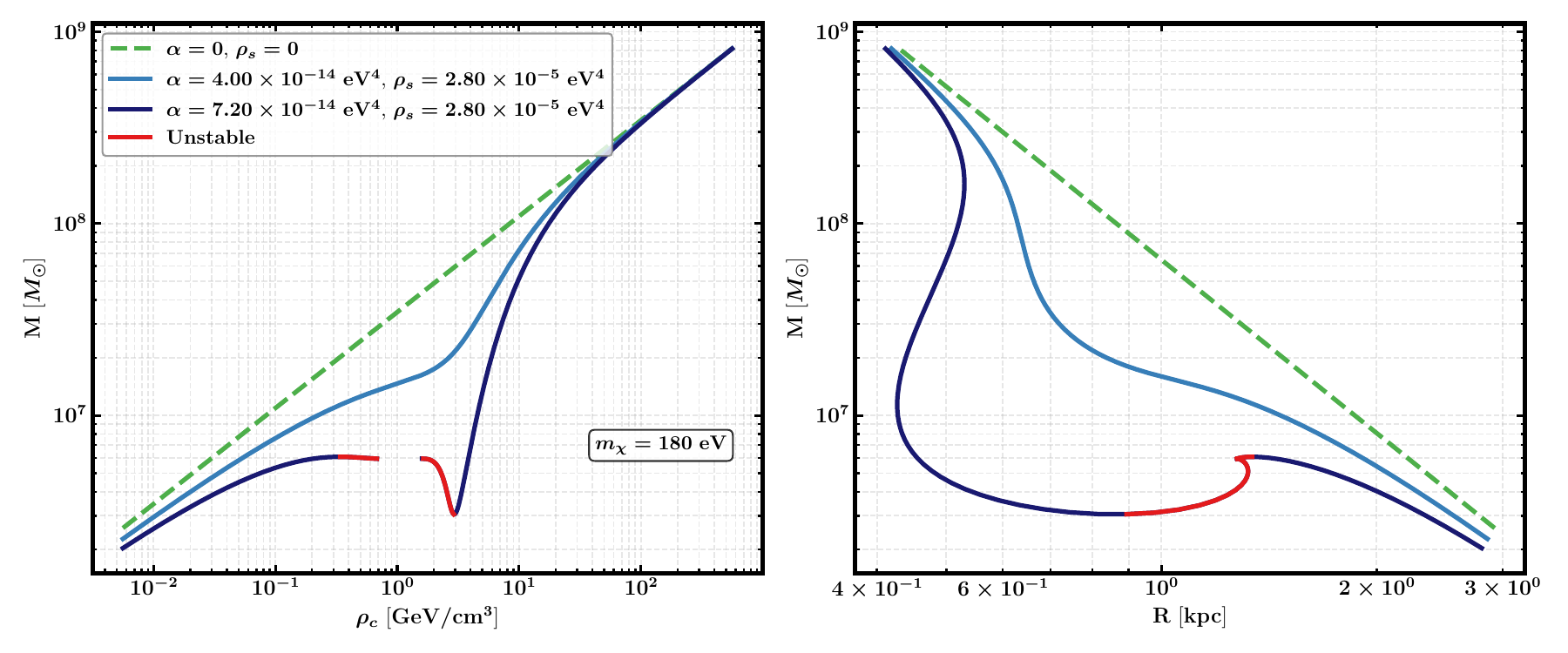}

    \caption{\justifying  Mass--radius curve of DM halos for fixed fermion mass $m_\chi = 180\,{\rm eV}$ and interaction scale density $\rho_s = 2.80\times 10^{-5}\,{\rm eV}^4$, with varying interaction strength $\alpha$. The left panel shows the total halo mass as a function of the central density $\rho_c$, while the right panel shows the corresponding mass--radius relation. The green curve denotes the non-interacting degenerate Fermi-gas, corresponding to $\alpha=0$, while the blue and purple curves show interacting-EoS sequences. Attractive interaction suppresses the effective sound speed and can generate a mechanically unstable interval in the equation of state. After the Maxwell construction, central densities lying inside the constant-pressure coexistence region are excluded, producing the gap visible in the $M(\rho_c)$ curve. In the mass--radius plane, the same excluded interval is mapped into the cusp-like feature of the corresponding curve. Red points on the purple sequence denote configurations satisfying $dM/d\rho_c<0$, which lie beyond the turning point of the equilibrium configurations and are expected to be dynamically unstable.}
    \label{fig:mass-radius_stability}
\end{figure*}

The bottom-right panel of fig.~\ref{fig:model_comparison} shows the corresponding line-of-sight velocity-dispersion profiles for the Carina dwarf spheroidal. These curves are obtained by inserting the hydrostatic solutions into the spherical Jeans calculation in Eq.~\eqref{eq:vel_disp}. The qualitative effect is clear: for fixed $m_\chi$ and $\rho_c$, the attractive interaction reduces the effective core size and lowers the enclosed mass over the radial range probed by the stellar kinematic tracers. Depending on the central density, this can either improve or worsen the agreement with the observed velocity-dispersion data. For example, an interaction that reduces an overly extended non-interacting core can bring the prediction closer to the data. On the other hand, if the interaction makes the halo too compact, the predicted velocity dispersion can fall below the observed profile, as illustrated by the blue solid curve in fig.~\ref{fig:model_comparison}, which gives a poorer fit than the corresponding dashed non-interacting curve.

This comparison shows that the interacting EoS does not simply produce a uniform improvement over the free Fermi-gas model. Rather, its effect depends on where the hydrostatic solution lies relative to the softened or unstable region of the equation of state. Since the non-interacting degenerate Fermi-gas model already provides acceptable fits to several dwarf-galaxy data sets~\cite{Domcke:2014kla}, the role of the present analysis is not only to test whether interactions can improve the fit, but also to determine which regions of the interaction parameter space $(\alpha,\rho_s)$ are excluded because they lead to halos that are too compact, too diffuse, or mechanically unstable.

\subsubsection{Varying DM mass}

We now allow the fermion mass to vary. This introduces an additional degeneracy between the DM mass scale and the strength of the attractive interaction. In the non-interacting degenerate Fermi-gas model, lowering $m_\chi$ increases the degeneracy pressure at fixed density and therefore produces a larger core~\cite{Domcke:2014kla}. By contrast, the attractive interaction considered here suppresses $c_s^2$ over a finite density interval and steepens the density profile whenever the hydrostatic solution samples this region. The two effects can therefore partially compensate each other: a lighter fermion, which would otherwise lead to a core that is too extended, may remain compatible with the observed stellar kinematics if the interaction reduces the core size sufficiently. This reduction may occur smoothly, through the suppression of $c_s^2$, or discontinuously, through the Maxwell construction when the solution crosses a mechanically unstable branch.

An example of this compensating effect is shown in fig.~(\ref{fig:mass_degen}) for the Carina dwarf spheroidal. In the absence of interactions, decreasing the fermion mass from $m_\chi = 200\,{\rm eV}$ to $m_\chi = 150\,{\rm eV}$ increases the core size and worsens the agreement with the observed velocity-dispersion profile. However, when attractive interactions are considered, the $m_\chi = 150\,{\rm eV}$ profile can be made significantly more compact. As shown by the solid blue curve, the resulting line-of-sight velocity dispersion can closely mimic the non-interacting $m_\chi = 200\,{\rm eV}$ case. This demonstrates that dwarf-galaxy kinematics constrain not only the fermion mass, but also the interaction parameters that shape the EoS. We quantify this degeneracy, and the corresponding limits on $(\alpha,\rho_s)$, in Sec.~\ref{sec:results} using a Markov Chain Monte Carlo analysis.

\subsection{Mass-Radius diagram}\label{sec:mass-radius}

We now examine the global mass--radius structure of the hydrostatic solutions. Figure~(\ref{fig:mass-radius_stability}) shows the total DM halo mass as a function of the central density $\rho_c$ in the left panel, and the corresponding mass--radius relation in the right panel. The fermion mass is fixed to $m_\chi = 180\,{\rm eV}$, while the interaction scale is kept at $\rho_s = 2.80\times 10^{-5}\,{\rm eV}^4$.
Different curves correspond to different values of the interaction strength $\alpha$. The green curve shows the standard non-interacting degenerate Fermi-gas sequence, while the blue and purple curves show the corresponding sequences for the interacting EoS for $\alpha = 4\times 10^{-14}\,{\rm eV^4}$ and $\alpha = 7.2\times 10^{-14}\,{\rm eV^4}$, respectively.

The non-interacting curve exhibits the well known behavior of a self-gravitating degenerate Fermi gas: increasing the central density increases the total mass and reduces the characteristic radius, producing the familiar monotonic mass--radius trend. Once attractive interactions are included, this structure can be modified substantially. The interaction suppresses the effective sound speed over a finite range of densities and can also generate a mechanically unstable branch in the equation of state. When such an unstable branch appears, it is replaced by the Maxwell-constructed constant-pressure segment, as described in appendix~\ref{app:maxwell_construction}. Central densities that lie inside this coexistence region do not correspond to regular hydrostatic solutions. This is why a gap appears in the $M(\rho_c)$ curve in the left panel of fig.~(\ref{fig:mass-radius_stability}).

The same feature appears differently in the mass--radius plane. Since the hydrostatic solution jumps between the two stable branches of the equation of state at fixed pressure, the excluded interval in central density is mapped into a sharp turn, or cusp-like feature, in the $M(R)$ curve. Thus, the gap in the left panel and the cusp in the right panel are two representations of the same underlying physics: a phase transition induced by the interacting equation of state. 

The red points on the purple curve mark configurations for which
\[
\frac{dM}{d\rho_c}<0 .
\]
These solutions lie beyond the turning point of the equilibrium sequence and are expected to be dynamically unstable according to the usual stability criterion for self-gravitating fluids~\cite{Shapiro:1983du}. The onset of this unstable branch shows that the attractive interaction does not only change the size of individual DM halo cores; it also modifies the stability structure of the full family of hydrostatic configurations.

This observation is important for interpreting fits to dwarf-galaxy stellar kinematics. A given interacting equation of state may produce a density profile that improves the agreement with the observed velocity-dispersion data, but such a solution is physically viable only if it belongs to a stable branch of the hydrostatic sequence. The mass--radius diagram therefore provides a useful complementary diagnostic: it identifies which regions of parameter space lead to acceptable equilibrium halos and which regions should be excluded because they correspond to unstable configurations or to central densities inside the Maxwell coexistence region. The qualitative behaviour found here for the phenomenological EoS in Eq.~\eqref{eq:eos_ph} is consistent with similar results obtained for microscopic attractive Yukawa models~\cite{Gresham:2018rqo}; see also appendix~\ref{app:UV_model}.

\section{MCMC setup and Results}\label{sec:results}

In this section, we constrain the parameters of the fermionic DM EoS by comparing the predicted line-of-sight velocity-dispersion profiles, $\sigma_{\rm los}(R)$, with stellar kinematic data from the eight classical Milky Way dwarf spheroidal galaxies: Carina, Draco, Fornax, Leo I, Leo II, Sculptor, Sextans, and Ursa Minor. We use the binned velocity-dispersion profiles compiled in Ref.~\cite{Domcke:2014kla}, which are obtained from line-of-sight velocity measurements of individual member stars by minimizing a negative log-likelihood in each radial bin~\cite{Walker:2005nt, Walker:2007ju, Walker:2009}.

Our aim is twofold. First, we determine whether the non-interacting degenerate Fermi-gas model already provides acceptable fits to the stellar kinematics. Second, we ask how well the intercating EoS can be tested with the same data. The latter question is particularly important because, as discussed in the previous section, attractive interactions can reduce the effective core size. 

We consider two models. The first is the non-interacting degenerate Fermi gas, with the EoS given by Eq.~\eqref{eq:eos_fermi}, with parameter vector
\begin{equation}
    \boldsymbol{\theta}_{\rm free} = (m_\chi,\rho_c,\beta),
\end{equation}
where $m_\chi$ is the fermion mass, $\rho_c$ is the central dark-matter density, and $\beta$ is the stellar velocity-anisotropy parameter introduced in eq.\eqref{eq:jeans}. The second is the interacting model defined by Eq.~\eqref{eq:eos_ph}, with $\alpha,\rho_s\neq 0$, and parameter vector
\begin{equation}
    \boldsymbol{\theta}_{\rm int}
    =
    (m_\chi,\alpha,\rho_s,\rho_c,\beta).
\end{equation}

For each galaxy, the two models are fitted independently. For a given point in parameter space, the predicted velocity-dispersion profile is obtained through the following sequence. First, the equation of state determines the sound speed, $c_s^2(\rho)$. This quantity enters the non-relativistic hydrostatic equations, Eqs.~\eqref{eq:continuity} and~\eqref{eq:equilibrium}. We then integrate these equations outward from the center with initial conditions $\rho(0)=\rho_c$ and $M(0)=0$,
to obtain the DM density profile $\rho(r)$ and the enclosed mass profile $M(r)$. In the non-interacting case, this procedure reduces to solving the Lane--Emden equation with polytropic index $n=3/2$.

For the interacting EoS, some parameter choices produce an interval in which $c_s^2<0$. Such a region is mechanically unstable and does not admit a continuous hydrostatic solution. In these cases, we attempt to implement the Maxwell construction described in appendix~\ref{app:maxwell_construction}. If a valid transition pressure exists, the hydrostatic solution is continued by jumping at fixed pressure from the high-density branch to the low-density branch, as described in Sec.~\ref{sec:hydro}. If no physically admissible Maxwell construction exists, the parameter point is rejected.

The resulting enclosed mass profile is then inserted into the Jeans expression for the projected velocity dispersion, Eq.~(\ref{eq:vel_disp}), together with a Plummer profile for the stellar tracer population. This gives the predicted line-of-sight velocity dispersion,
$\sigma_{\rm los}^{\rm pred}(R_i;\boldsymbol{\theta})$,
at the projected radii of the observed data points.

Assuming uncorrelated Gaussian errors, the log-likelihood for a single galaxy is
\begin{equation}
\label{eq:log_likelihood}
\ln \mathcal{L}(\boldsymbol{\theta}) = -\frac{1}{2} \sum_{i=1}^{N} \frac{\left[ \sigma_{\rm los}^{\rm pred}(R_i;\boldsymbol{\theta}) - \sigma_{\rm los}^{\rm obs}(R_i) \right]^2}{\delta\sigma_i^2},
\end{equation}
up to an additive constant. Here $N$ is the number of radial bins, $\sigma_{\rm los}^{\rm obs}(R_i)$ is the observed velocity dispersion in the $i$-th bin, and $\delta\sigma_i$ is the corresponding $1\sigma$ uncertainty.\\

\noindent\underline{\it Priors:} We adopt broad uniform priors for all model parameters. For the fermion mass and stellar anisotropy, we use
\begin{equation}
    m_\chi \in [5,500]~{\rm eV}, \qquad \beta \in [-1.5,1].
\end{equation}
The central density is sampled logarithmically with a flat prior
\begin{equation}
    \log_{10} \left(\frac{\rho_c}{{\rm kg\,m}^{-3}}\right) \in [-22,-19],
\end{equation}
corresponding to  $\rho_c \in [0.056,56.096]~{\rm GeV/cm}^{3}$.

For the interacting EoS, we have two additional parameters $\alpha, \rho_c$. For the interaction strength $\alpha$, we choose
\begin{equation}
    \log_{10}\left(\frac{\alpha}{{\rm eV}^{4}}\right)\in [-20,-9].
\end{equation}
The lower edge of this prior corresponds to an interaction pressure that is negligible compared with the Fermi degeneracy pressure across the parameter range considered. Thus, $\alpha=10^{-20}~{\rm eV}^4$ effectively reproduces the non-interacting limit. The upper edge is chosen conservatively: for $\alpha=10^{-9}~{\rm eV}^4$, the attractive interaction beats the degeneracy pressure and the hydrostatic equations typically admit no physical solution, even after attempting the Maxwell construction.  This way the sampled parameter space includes both the non-interacting regime and the region where strong attractive interactions are excluded either by the absence of stable cores or by the kinematic data.

We also sample the scale density $\rho_s$ logarithmically,
\begin{equation}
    \log_{10}\left(\frac{\rho_s}{{\rm kg\,m}^{-3}}\right)\in [-22,-19],
\end{equation}
corresponding to $\rho_s\in  [4.31\times 10^{-7},4.31\times 10^{-4}]~{\rm eV}^{4}.$

This range is chosen to match the central-density prior, so that the dipping feature in the EoS can occur within the density range probed by the dwarf-galaxy solutions. If $\rho_s$ is much larger than the relevant $\rho_c$, the degeneracy pressure dominates throughout the halo and the model reduces to the non-interacting case. If $\rho_s$ is much smaller than the relevant $\rho_c$, the interaction term has already saturated and its derivative is negligible, so $c_s^2$ again approaches the non-interacting value. In either limit, $\rho_s$ becomes difficult to constrain.

We sample the posterior using the affine-invariant ensemble sampler \texttt{emcee}~\cite{Foreman_Mackey_2013}. For each galaxy and each model, we use $40$ walkers and $4\times 10^4$ iterations, discarding the first $2\times 10^4$ samples as burn-in. We check convergence using the criterion recommended in the \texttt{emcee} documentation: the post-burn-in chain length must exceed $50\tau_{\rm max}$, where $\tau_{\rm max}$ is the largest integrated autocorrelation time among the sampled parameters. We also monitor the walker acceptance fractions. For the non-interacting model, the mean acceptance fraction lies in the range $0.6$--$0.65$, while for the interacting model it lies in the range $0.3$--$0.5$. Although the non-interacting acceptance fractions are somewhat high, the autocorrelation-time estimates as well as the posteriors converge satisfactorily for all galaxies.\\

\begin{table*}
\centering
\renewcommand{\arraystretch}{1.3}
\begin{tabular}{l l c c c c c c}
\hline\hline
Galaxy & Model
  & $m_f$ [eV]
  & $\log_{10}(\rho_c [\mathrm{GeV\,cm}^{-3}])$
  & $\beta$
  & $\log_{10}(\alpha\ [\mathrm{eV}^{4}])\big|_{95\%}$
  & $M_{\mathrm{dyn}}(r_{\text{half}})$
  & $M_{\mathrm{dyn}}(r_{\text{half}})$~\cite{Walker:2009} \\
 & & & & & & [$10^7\,M_\odot$] & [$10^7\,M_\odot$] \\
\hline
\multirow{2}{*}{Carina}
  & interacting & $\mathrm{171.89}^{+\mathrm{28.61}}_{-\mathrm{20.06}}$ & $\mathrm{0.31}^{+\mathrm{0.11}}_{-\mathrm{0.09}}$ & $\mathrm{0.33}^{+\mathrm{0.16}}_{-\mathrm{0.23}}$ & $\leq\mathrm{-12.98}$ & 0.29 & $\mathrm{0.4}^{+\mathrm{0.1}}_{-\mathrm{0.1}}$ \\
  & non-interacting & $\mathrm{172.93}^{+\mathrm{28.85}}_{-\mathrm{21.23}}$ & $\mathrm{0.31}^{+\mathrm{0.11}}_{-\mathrm{0.09}}$ & $\mathrm{0.33}^{+\mathrm{0.16}}_{-\mathrm{0.23}}$ & --- & 0.29 & \\
\hline
\multirow{2}{*}{Draco}
  & interacting & $\mathrm{151.88}^{+\mathrm{11.17}}_{-\mathrm{8.26}}$ & $\mathrm{0.69}^{+\mathrm{0.08}}_{-\mathrm{0.07}}$ & $\mathrm{0.16}^{+\mathrm{0.23}}_{-\mathrm{0.33}}$ & $\leq\mathrm{-13.31}$ & 0.39 & $\mathrm{0.6}^{+\mathrm{0.5}}_{-\mathrm{0.3}}$ \\
  & non-interacting & $\mathrm{151.88}^{+\mathrm{13.35}}_{-\mathrm{8.33}}$ & $\mathrm{0.69}^{+\mathrm{0.07}}_{-\mathrm{0.08}}$ & $\mathrm{0.17}^{+\mathrm{0.22}}_{-\mathrm{0.34}}$ & --- & 0.39 & \\
\hline
\multirow{2}{*}{Fornax}
  & interacting & $\mathrm{137.54}^{+\mathrm{6.58}}_{-\mathrm{6.99}}$ & $\mathrm{0.24}^{+\mathrm{0.06}}_{-\mathrm{0.06}}$ & $\mathrm{0.11}^{+\mathrm{0.09}}_{-\mathrm{0.10}}$ & $\leq\mathrm{-12.81}$ & 4.04 & $\mathrm{4.3}^{+\mathrm{0.6}}_{-\mathrm{0.7}}$ \\
  & non-interacting & $\mathrm{138.45}^{+\mathrm{6.30}}_{-\mathrm{6.20}}$ & $\mathrm{0.24}^{+\mathrm{0.06}}_{-\mathrm{0.06}}$ & $\mathrm{0.11}^{+\mathrm{0.09}}_{-\mathrm{0.10}}$ & --- & 4.04 & \\
\hline
\multirow{2}{*}{Leo~I}
  & interacting & $\mathrm{166.31}^{+\mathrm{22.56}}_{-\mathrm{17.88}}$ & $\mathrm{0.73}^{+\mathrm{0.12}}_{-\mathrm{0.10}}$ & $\mathrm{0.28}^{+\mathrm{0.23}}_{-\mathrm{0.35}}$ & $\leq\mathrm{-13.11}$ & 0.80 & $\mathrm{1.0}^{+\mathrm{0.6}}_{-\mathrm{0.4}}$ \\
  & non-interacting & $\mathrm{166.77}^{+\mathrm{22.72}}_{-\mathrm{17.88}}$ & $\mathrm{0.73}^{+\mathrm{0.12}}_{-\mathrm{0.10}}$ & $\mathrm{0.28}^{+\mathrm{0.23}}_{-\mathrm{0.35}}$ & --- & 0.79 & \\
\hline
\multirow{2}{*}{Leo~II}
  & interacting & $\mathrm{292.99}^{+\mathrm{61.54}}_{-\mathrm{50.57}}$ & $\mathrm{1.19}^{+\mathrm{0.27}}_{-\mathrm{0.21}}$ & $\mathrm{0.24}^{+\mathrm{0.38}}_{-\mathrm{0.73}}$ & $\leq\mathrm{-13.80}$ & 0.45 & $\mathrm{0.5}^{+\mathrm{0.2}}_{-\mathrm{0.3}}$ \\
  & non-interacting & $\mathrm{301.84}^{+\mathrm{73.33}}_{-\mathrm{56.61}}$ & $\mathrm{1.23}^{+\mathrm{0.33}}_{-\mathrm{0.23}}$ & $\mathrm{0.18}^{+\mathrm{0.43}}_{-\mathrm{0.84}}$ & --- & 0.47 & \\
\hline
\multirow{2}{*}{Sculptor}
  & interacting & $\mathrm{158.58}^{+\mathrm{12.73}}_{-\mathrm{12.48}}$ & $\mathrm{0.61}^{+\mathrm{0.06}}_{-\mathrm{0.06}}$ & $\mathrm{0.23}^{+\mathrm{0.11}}_{-\mathrm{0.13}}$ & $\leq\mathrm{-12.75}$ & 0.71 & $\mathrm{1.0}^{+\mathrm{0.3}}_{-\mathrm{0.3}}$ \\
  & non-interacting & $\mathrm{159.76}^{+\mathrm{12.66}}_{-\mathrm{11.92}}$ & $\mathrm{0.61}^{+\mathrm{0.06}}_{-\mathrm{0.06}}$ & $\mathrm{0.23}^{+\mathrm{0.11}}_{-\mathrm{0.13}}$ & --- & 0.71 & \\
\hline
\multirow{2}{*}{Sextans}
  & interacting & $\mathrm{208.96}^{+\mathrm{23.84}}_{-\mathrm{36.40}}$ & $\mathrm{0.20}^{+\mathrm{0.20}}_{-\mathrm{0.28}}$ & $\mathrm{-0.35}^{+\mathrm{0.34}}_{-\mathrm{0.42}}$ & $\leq\mathrm{-12.93}$ & 1.87 & $\mathrm{1.6}^{+\mathrm{0.4}}_{-\mathrm{0.4}}$ \\
  & non-interacting & $\mathrm{212.08}^{+\mathrm{22.68}}_{-\mathrm{35.48}}$ & $\mathrm{0.21}^{+\mathrm{0.19}}_{-\mathrm{0.28}}$ & $\mathrm{0.35}^{+\mathrm{0.34}}_{-\mathrm{0.40}}$ & --- & 1.87 & \\
\hline
\multirow{2}{*}{UMi}
  & interacting & $\mathrm{189.14}^{+\mathrm{45.99}}_{-\mathrm{34.16}}$ & $\mathrm{0.73}^{+\mathrm{0.23}}_{-\mathrm{0.16}}$ & $\mathrm{-0.10}^{+\mathrm{0.30}}_{-\mathrm{0.49}}$ & $\leq\mathrm{-12.96}$ & 1.06 & $\mathrm{1.3}^{+\mathrm{0.3}}_{-\mathrm{0.5}}$ \\
  & non-interacting & $\mathrm{192.63}^{+\mathrm{46.15}}_{-\mathrm{34.67}}$ & $\mathrm{0.74}^{+\mathrm{0.23}}_{-\mathrm{0.16}}$ & $\mathrm{0.10}^{+\mathrm{0.5}}_{-\mathrm{0.5}}$ & --- & 1.07 & \\
\hline\hline
\end{tabular}
\caption{\justifying Posterior summary for the interacting and non-interacting models for the eight classical dwarf spheroidals. 
We report the median and 68\% credible intervals for $m_f$, $\log_{10}\rho_c$, and $\beta$, the 95\% upper limit on $\log_{10}\alpha$, the median of the mass contained with the half-light radius $M_{\mathrm{dyn}}(r_{\text{half}})$, and the inferred mass estimate $M_{\mathrm{dyn}}(r_{\text{half}})$ from \cite{Walker:2009}}
\label{tab:median_vals}
\end{table*}

\noindent\underline{\it Physical consistency cuts:} During sampling, we reject parameter points for which the hydrostatic integration does not yield a physical halo profile, or for which the interacting equation of state contains a mechanically unstable region but the Maxwell construction does not produce a valid transition pressure. We also reject central densities that fall within the Maxwell-constructed constant pressure region. 

After sampling, we impose two additional physical cuts on the posterior. The first is the Tremaine--Gunn condition. For each posterior sample, we compute the characteristic Fermi velocity
\begin{equation}
    v_F=\left(\frac{3\bar{\rho}}{8\pi m_\chi^4}\right)^{1/3},
\end{equation}
where $ \bar{\rho}=3M/(4\pi R^3)$, is the mean density of the hydrostatic solution, with $M$ and $R$ denoting its total mass and radius. We require
\begin{equation}
    v_F \leq v_{\rm esc}.
\end{equation}
Following Ref.~\cite{Domcke:2014kla}, we estimate the escape velocity from the observed global velocity dispersion as
\begin{equation}
    v_{\rm esc}=\sqrt{6}\,\sigma_{\rm global}.
\end{equation}
Samples that violate this condition are discarded. The impact of this cut varies substantially across the dwarf galaxies: for Fornax, Sextans and UMi this removes $<5\%$ of the samples, while for Carina, Leo I and Leo II it removes $\sim20\%$ of the samples. For Draco it removes approximately $90\%$. Thus, for Draco, the Tremaine--Gunn condition provides the dominant lower constraint on the fermion mass.

The second cut requires that the DM halo extend at least to the observed stellar half-light radius,
\begin{equation}
    R_{\rm tot} > r_{\text{half}}.
\end{equation}
This removes solutions that are too compact to contain the observed stellar distribution. The effect is most pronounced for Sextans, which has the largest half-light radius among the galaxies considered. For Sextans, this condition removes approximately $37\%$ of the posterior samples, preferentially excluding compact-core solutions. For all other galaxies, this imposes negligible or zero cuts to the posterior. 

In principle, both cuts could be imposed directly during sampling. In practice, doing so significantly lowers the acceptance fraction because the combination of broad priors and hard physical boundaries causes many proposed steps to land in excluded regions. We therefore impose these cuts in post-processing. As a validation, we repeated the Sextans analysis with narrower uniform priors chosen from the $3\sigma$ credible intervals of the uncut posterior for the parameters $m_\chi$, $\log_{10}\rho_c$, and $\beta$, while enforcing the viability cuts during sampling. The priors on $\log_{10}\alpha$ and $\log_{10}\rho_s$ were left unchanged. The resulting posteriors were consistent with the post-processed results, confirming that the procedure gives stable constraints.\\

\noindent\underline{\it Further comments:} We check whether the posterior samples correspond to dynamically stable hydrostatic configurations. As discussed in Sec.~\ref{sec:mass-radius}, configurations on branches satisfying $\frac{\mathrm{d}M}{\mathrm{d}\rho_c}<0$, are expected to be dynamically unstable. A direct implementation of this criterion for every posterior sample would require solving the hydrostatic equations at neighbouring central densities, for example at $\rho_c(1\pm 10^{-3})$, in order to estimate the local derivative $\mathrm{d}M/\mathrm{d}\rho_c$. This would be computationally expensive, increasing the MCMC runtime by roughly a factor of $2$--$3$.

We therefore perform this stability check as a post-processing step. Since the data primarily impose an upper bound on the interaction strength, most posterior samples after the cuts discussed above lie close to the non-interacting limit, where the hydrostatic sequence is stable. This behaviour is illustrated by the blue curve in fig.~\ref{fig:mass-radius_stability}. Possible instabilities are expected only in the large-$\alpha$ tail, where the interacting EoS departs significantly from the free Fermi-gas case.

For this reason, after imposing the previous cuts on the posterior, we examine $5000$ samples drawn from the neighbourhood of the $95\%$ upper limit on $\log_{10}\alpha$, corresponding to the $90$--$99$ percentile tail of the posterior. In this region the interaction is large enough for stability issues, if present, to become relevant. Among the dwarf galaxies considered, Carina gives the largest number of unstable configurations: $99$ out of $5000$ samples, corresponding to approximately $2\%$ of this high-$\alpha$ subset. Sculptor, Sextans, and Ursa Minor contain only a small number of unstable samples, with $16/5000$, $24/5000$, and $20/5000$ respectively, all below $0.5\%$. The remaining galaxies show no unstable configurations in this region.

To estimate the impact of this stability requirement on the inferred exclusion bound, we focus on the most conservative case, namely Carina. We thin the post-processed chain by a factor of $50$, retaining every $50$th sample and reducing the chain to approximately $1.3\times 10^4$ samples. We then remove all configurations satisfying $\mathrm{d}M/\mathrm{d}\rho_c<0$ and recompute the $95$th percentile upper limit on $\log_{10}\alpha$. The shift in the resulting bound is only about $0.03$ dex. Thus, even in the worst-case case of Carina, removing dynamically unstable configurations has only a minor effect on the quoted upper limit.

\subsection{Results}

\begin{figure*}[htb!]
    \centering
    \includegraphics[width=0.47\linewidth]{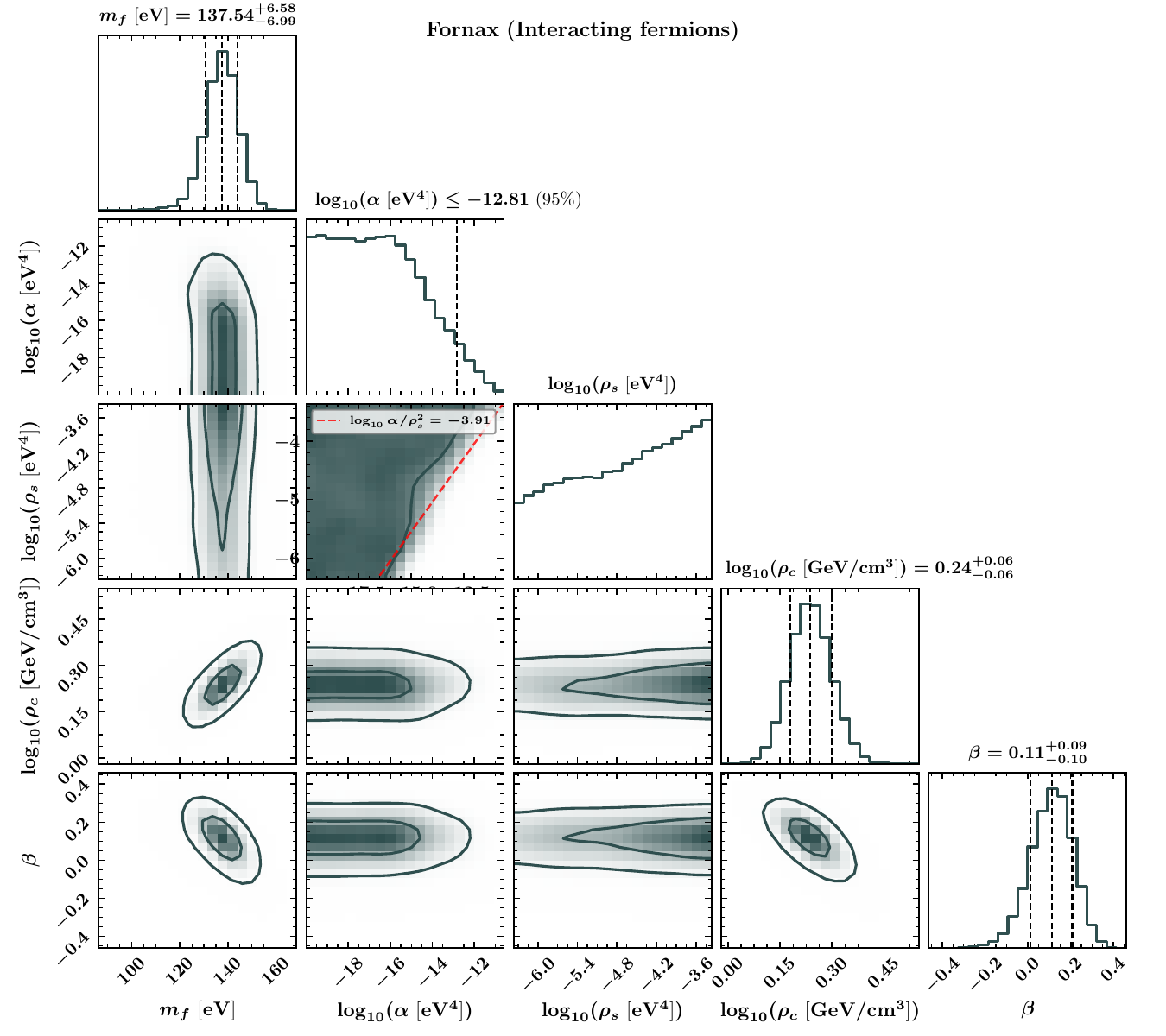}
    \includegraphics[width=0.47\linewidth]{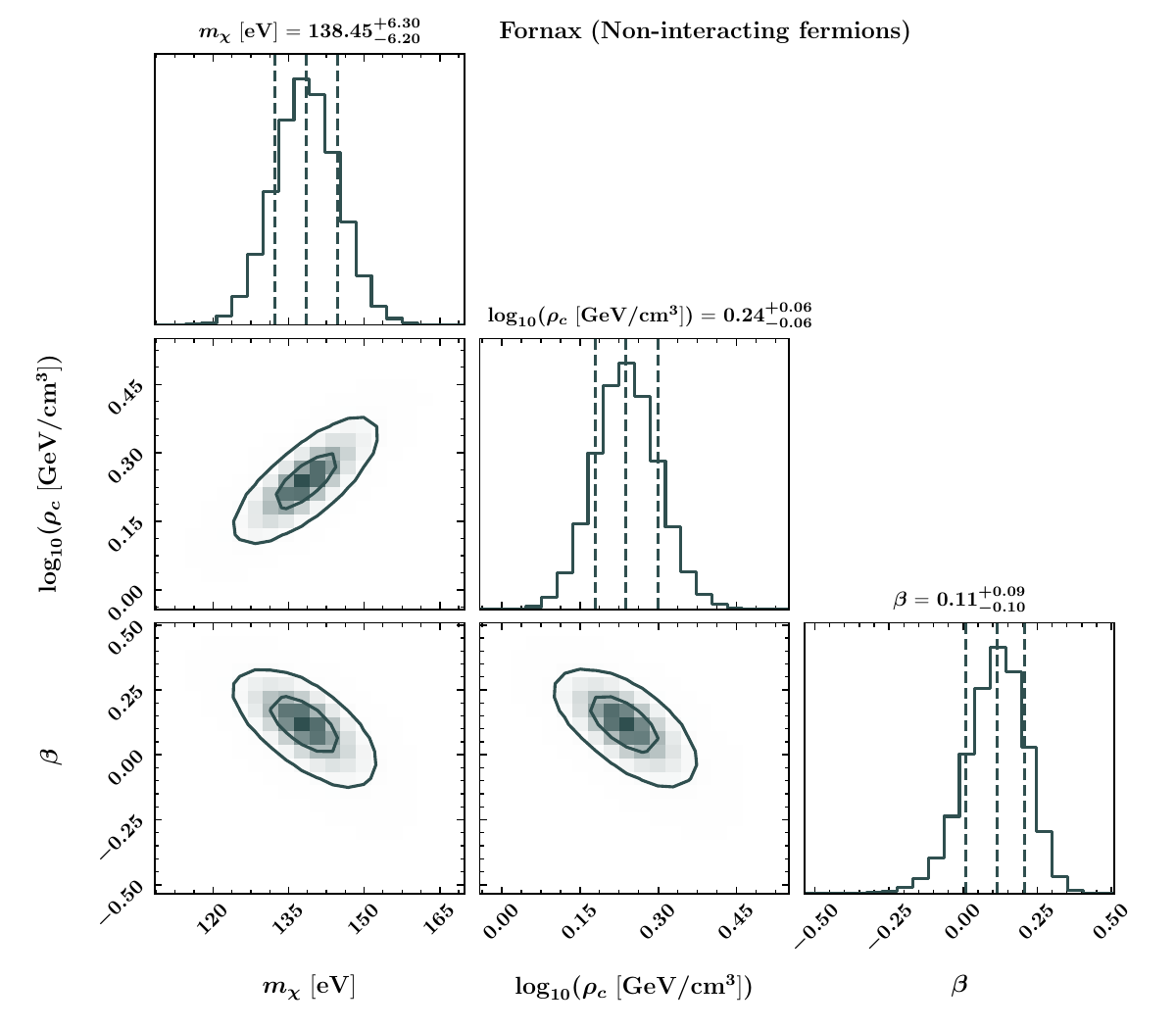}\\
    \includegraphics[width=0.47\linewidth]{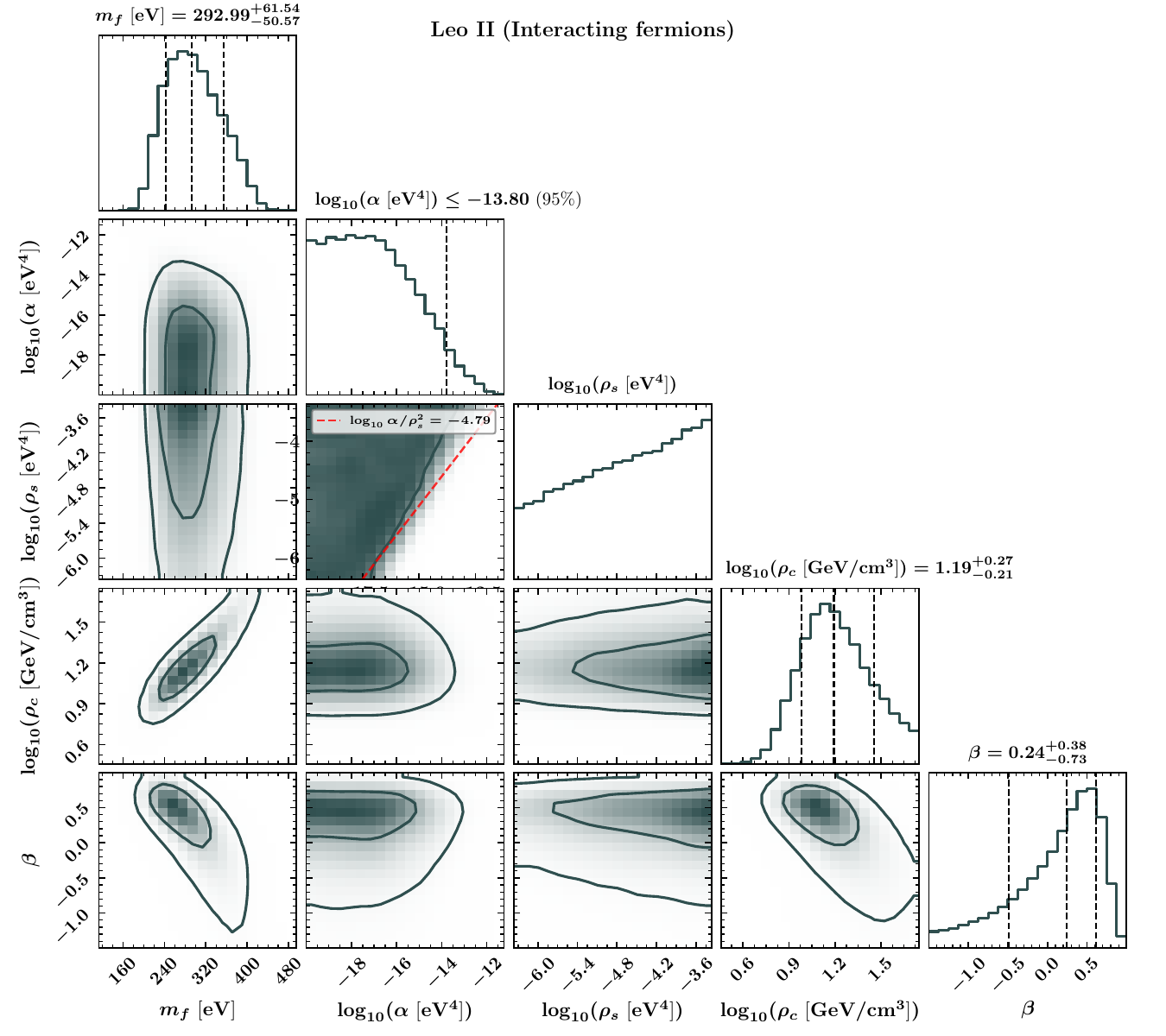}
    \includegraphics[width=0.47\linewidth]{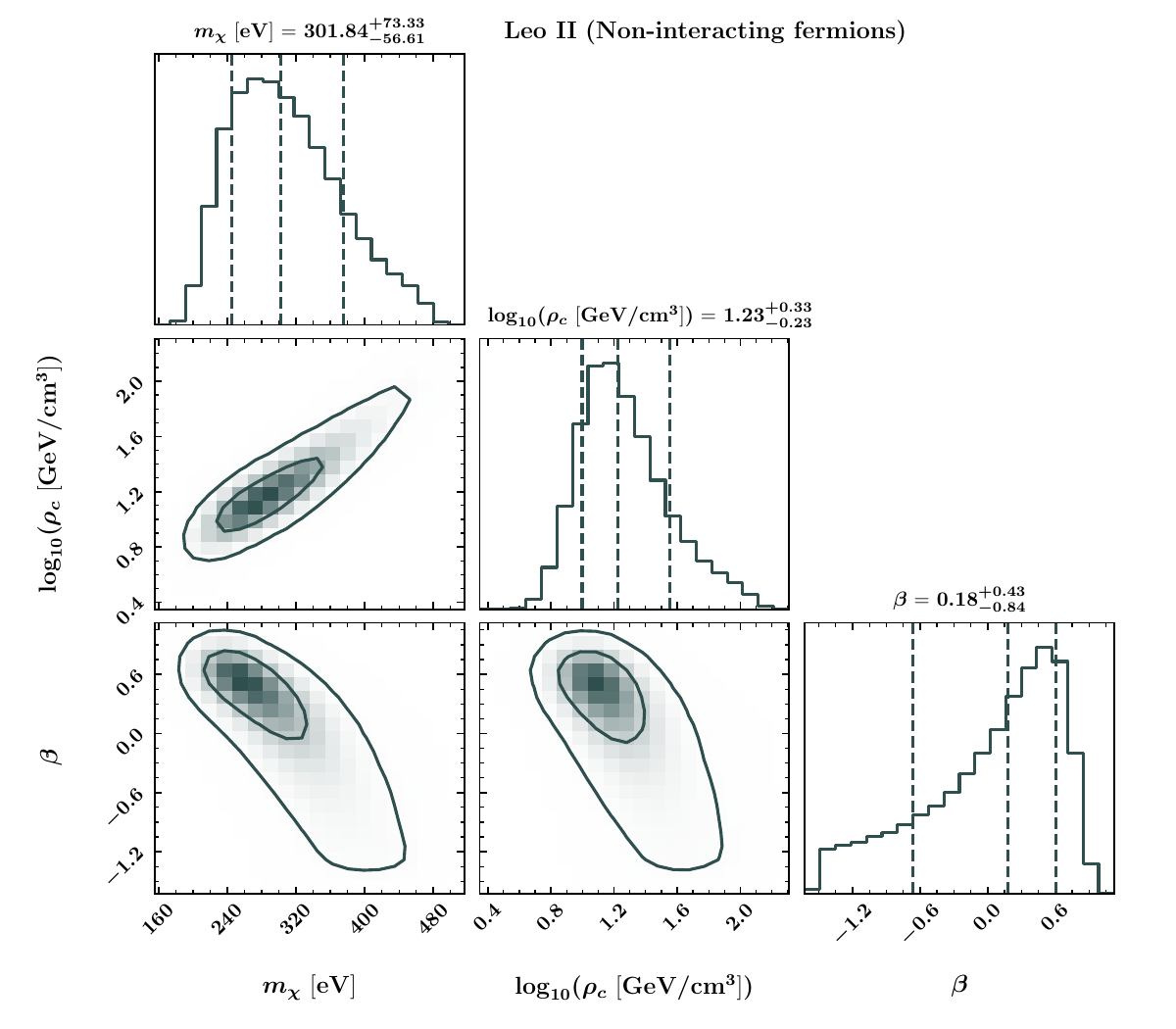}
    \caption{\justifying Posterior distributions from single-galaxy MCMC fits to Fornax (top) and Leo II (bottom). Left panels show the full five-parameter interacting model ($m_\chi,\,\log_{10}\alpha,\,\log_{10}\rho_s,\,\log_{10}\rho_c,\,\beta$) while the right panels show the three-parameter non-interacting (free Fermi gas) baseline ($m_\chi,\,\log_{10}\rho_c,\,\beta$). 
    Contours in the 2D panels enclose the $1\sigma$ and $2\sigma$ credible regions, except the $\log_{10}\rho_s-\log_{10}\alpha$ contour where we only show the $2\sigma$.
    Dashed lines in the 1D marginals indicate the median and 68\% credible interval, except for $\log_{10}\alpha$ where we show the one-sided 95\% upper limit. 
    The fermion mass $m_\chi$ is well constrained in both models and consistent between the interacting and non-interacting fits for each galaxy. The posteriors for parameters characterizing the interactions indicate that small $\alpha$ and large $\rho_s$ remain consistent with data, signaling that one can only obtain upper limits for $\alpha$ given the priors.}
    \label{fig:fornax_leoII_post}
\end{figure*}

The posterior summaries for the eight classical dwarf spheroidals are reported in table~\ref{tab:median_vals}. Representative posterior distributions are shown in fig.~(\ref{fig:fornax_leoII_post}) for Fornax and Leo~II, comparing the full interacting EoS with the non-interacting degenerate Fermi-gas baseline. Similar results for the remaining dwarf spheroidals are presented in appendix~\ref{app:mcmc_results}. These two galaxies are useful examples: Fornax has one of the best measured velocity-dispersion profiles and gives a relatively tight constraint on the halo parameters, while Leo~II yields the strongest upper bound on the interaction strength $\alpha$ among the galaxies considered. We discuss the results parameter by parameter.\\

\noindent\underline{\it Fermion mass:} The fermion mass $m_\chi$ is well constrained by the velocity-dispersion data for both the interacting and non-interacting EoS. Across all galaxies, the preferred values lie in the range
\begin{equation}
    m_\chi \sim 140\text{--}300~{\rm eV},
\end{equation}
with the precise value depending on the galaxy. The posterior medians obtained in the interacting model are very close to those obtained in the non-interacting model. This is visible in table~\ref{tab:median_vals} and explicitly in Fig.~\ref{fig:fornax_leoII_post}, where the $m_\chi$ marginals for the interacting and free Fermi-gas fits overlap strongly for both Fornax and Leo~II. 

This agreement indicates that the data primarily constraints the extent of the halo through the fermion degeneracy pressure. Allowing for attractive interactions does not significantly shift the preferred fermion mass, because the data do not allow a large departure from the non-interacting equation of state.
It is also worth noting that for both the non-interacting as well as interacting EoS, no single value of $m_\chi$ is consistent with all LOS data considered in this work within $2\sigma$; this is primarily because of the large fermion mass preferred by the more compact Leo II dwarf.
\\

\noindent \underline{\it Central density:} The central density $\rho_c$ is also consistently determined in the two models. For each galaxy, the posterior for $\log_{10}\rho_c$ in the interacting model is nearly identical to that in the non-interacting model. This again shows that the observed line-of-sight velocity-dispersion profile fixes the characteristic halo density rather robustly.

There is, however, a galaxy-to-galaxy variation in the preferred central density. More compact systems, such as Leo~II, prefer larger values of $\rho_c$, while more extended systems, such as Fornax and Sextans, prefer slightly lower central densities. At fixed fermion mass, increasing $\rho_c$ increases the central gravitational potential and alters the scale radius over which the density profile falls.\\

\noindent\underline{\it Velocity anisotropy:} The stellar anisotropy parameter $\beta$ remains only moderately constrained. For most galaxies, the posterior permits a broad range of mildly radial, isotropic, and mildly tangential stellar orbits. This reflects the usual degeneracy between the DM mass profile and the stellar orbital anisotropy in Jeans analyses~\cite{Walker:2009, Domcke:2014kla}.

Importantly, the consideration of interacting EoS does not substantially change the preferred values or credible intervals for $\beta$. The $\beta$ posteriors in the interacting and non-interacting fits are very similar, as illustrated in Fig.~\ref{fig:fornax_leoII_post}. Thus, the interaction parameters do not appear to be absorbing the role of stellar anisotropy (as physically expected); rather, the data constrain the halo profile in a way that remains close to the free Fermi-gas solution.\\

\noindent\underline{\it Interaction strength:} The MCMC analysis yields upper limits on $\alpha$. For the galaxies considered here, the typical bound is
\begin{equation}
    \log_{10}\left(\frac{\alpha}{{\rm eV}^{4}}\right)\lesssim-13~,
\end{equation}
with mild variation between systems. The strongest bound is obtained for Leo~II,
\begin{equation}
    \log_{10}\left(\frac{\alpha}{{\rm eV}^{4}}\right)\leq -13.80\qquad(95\%~{\rm C.L.}),
\end{equation}
while the bounds from the remaining dwarfs are typically clustered around $\log_{10}(\alpha/{\rm eV}^4)\sim -13$.

The physical reason for this upper limit is straightforward. A larger attractive interaction suppresses the sound speed over the density range relevant for the halo, steepening the density profile and reducing the effective core size. If the interaction is too strong, the resulting mass profile produces a line-of-sight velocity dispersion incompatible with the observed stellar kinematics, or the equation of state fails to admit a stable hydrostatic solution. The data therefore prefer the region in which the attractive correction is small enough that the model remains close to the non-interacting Fermi gas.

This behaviour is seen in the left panels of Fig.~\ref{fig:fornax_leoII_post}. The posterior extends toward small $\alpha$, where the model smoothly approaches the non-interacting limit. By contrast, large values of $\alpha$ are excluded because they lead either to excessive modification of the DM halo profile or to unstable configurations.\\

\noindent\underline{\it Interaction scale density:} The density scale $\rho_s$ controls the values of density where interactions become important. Consequently, its posterior must be interpreted together with $\rho_c$. If $\rho_s$ lies well above the densities sampled by the halo, the interaction term is negligible over the relevant region and the model reduces to the non-interacting case. If $\rho_s$ lies well below the relevant halo densities, the interaction term has already saturated and its derivative is small, so the EoS again approaches the non-interacting case. The interaction is most strongly constrained when $\rho_s$ lies close to $\rho_c$. For this reason, the posterior generally allows large values of $\rho_s$ together with small values of $\alpha$, which will correspond to an effectively free Fermi gas. 

Equivalently, the data limit the combination $\alpha/\rho_s^2$ and not $\rho_s$ independently. If $\rho_s$ is increased at fixed $\alpha$, the dip in the EoS is pushed to larger density. Conversely, at fixed $\rho_s$, increasing $\alpha$ deepens the dip and makes the halo more compact. The posterior correlations in the $\alpha$--$\rho_s$ plane therefore reflect the fact that models with comparable values of $\alpha/\rho_s^2$ can produce similar departures from the non-interacting Fermi-gas profile, provided that $\rho_s$ lies within the values of the central density. The corner plots in Fig.~\ref{fig:fornax_leoII_post} show this structure clearly (note the the red dashed curve, which shows $\alpha/\rho_s^2 = \mathrm{const}$ evaluated at the 95th percentile of the derived posterior on $\alpha/\rho_s^2$, which the $2\sigma$ contour approximately follows). The posterior volume remains open toward the non-interacting regime, while the regions where $\alpha$ is large and $\rho_s$ places the dip in the EoS within the halo are excluded.\\

\noindent\underline{\it Comparison with dynamical masses:} As an additional check, table~\ref{tab:median_vals} reports the median valur of dynamical mass enclosed within the half-light radius, $M_{\rm dyn}(r_{\text{half}})$. These values are broadly consistent with $M_{\rm dyn}(r_{\rm half})$ from Ref.~\cite{Walker:2009}. This agreement shows that the fitted solutions reproduces the shape of $\sigma_{\rm los}(R)$, as well as characteristic enclosed masses on the scales where the stellar data are most constraining.\\

Overall, the MCMC analysis shows that the present stellar kinematic data do not strongly constrain attractive self-interactions in the fermionic DM EoS. The non-interacting degenerate Fermi-gas model already describes the data well for the classical dwarf spheroidals considered in this work.
The interacting model is therefore constrained through the absence of large deviations from this baseline: sufficiently strong attractive interactions would make the halos too compact, alter the projected velocity-dispersion profiles, or eliminate stable hydrostatic solutions altogether. Implications of limits on $\alpha, \rho_s$, for a realistic particle physics model, such as DM with Yukawa interactions, are discussed in appendix~\ref{app:UV_model}.

\section{Conclusions}\label{sec:conclusions}

We have studied whether the stellar kinematics of dwarf spheroidal galaxies can probe finite-density interactions in degenerate fermionic DM scenarios. The broader motivation is straightforward: if DM is made of degenerate fermions, on dwarf-galaxy scales, its macroscopic halo profile is determined by its EoS. Interactions in the dark sector can therefore modify the pressure support, the compressibility, and even the stability structure of the halo. These effects may leave observable imprints that can be tested with stellar kinematic data.

Our work builds on earlier studies of fermionic dark matter in dwarf spheroidals, which have mainly used phase-space arguments, Jeans analyses, or assumed halo profiles to constrain the fermion mass. For example, Refs.~\cite{Boyarsky:2008ju,DiPaolo:2017geq,Domcke:2014kla,Alvey:2020xsk,Savchenko:2019qnn} have shown that dwarf spheroidals provide powerful tests of fermionic DM because of their large inferred phase-space densities and low astrophysical backgrounds. These studies typically constrain the allowed degenerate DM fermion mass, while sometimes treating the microscopic DM physics indirectly. Our approach is complementary. Instead of starting from a phenomenological halo profile and asking what particle masses or cross sections are allowed, we start from the DM EoS itself and solve for the corresponding hydrostatic configuration. 

We considered two representative equations of state: the standard non-interacting degenerate Fermi gas, given by Eq.~\eqref{eq:eos_fermi}, and an interacting degenerate fermionic DM motivated by mean-field finite-density systems, given by Eq.~\eqref{eq:eos_ph}. The interacting EoS accounts for attractive interactions (a possible particle physics model is discussed in appendix~\ref{app:UV_model}).  Solving the non-relativistic hydrostatic equations, we found that attractive interactions generically steepen the density profile whenever the solution samples the region of suppressed sound speed. At fixed fermion mass and central density, this reduces the effective core size and lowers the enclosed mass over the radii relevant for stellar tracers. This effect is not universal in its observational impact: in some cases it improves the agreement with velocity-dispersion data by reducing an overly extended core, while in others it makes the halo too compact and worsens the fit. 

We confronted these models with line-of-sight velocity-dispersion data for the classical Milky Way dwarf spheroidals Carina, Draco, Fornax, Leo~I, Leo~II, Sculptor, Sextans, and Ursa Minor. For each galaxy, we performed independent MCMC fits using both the non-interacting and interacting equations of state. The preferred fermion masses are typically in the range
\[
m_\chi \sim 100\text{--}300\,{\rm eV}~.
\]
The inferred values of $m_\chi$, $\rho_c$, and the stellar anisotropy parameter $\beta$ are broadly similar in the two models. Present stellar kinematic data therefore only place upper limits on the strength of attractive interactions and their scale density by excluding regions in which the halos become too compact, distort the line-of-sight velocity-dispersion profile, or correspond to dynamically unstable hydrostatic configurations. For the phenomenological EoS considered here, the resulting limits are typically of order
\[
\alpha \lesssim 10^{-13}\,{\rm eV}^4,
\]
with mild galaxy-to-galaxy variation.

Looking ahead, several improvements would sharpen these conclusions. On the observational side, larger spectroscopic samples, better membership determination at large radii, and Gaia-assisted proper-motion information can reduce uncertainties in the stellar velocity-dispersion profiles; for classical dwarf spheroidal galaxies as well as Ultra-faint dwarfs.
Higher-order velocity moments and multiple stellar populations could also help break the well-known degeneracy between the DM profile and stellar anisotropy. On the theoretical side, it would be useful to perform a fit across multiple galaxies, allow for more flexible anisotropy profiles, and connect the phenomenological EoS more explicitly to particle physics models. Put together it may provide a more stringent test of interacting fermionic degenerate DM.

\section*{Acknowledgement}
R.G. acknowledges support from grant number  ANRF/ECRG/2025/000418/PMS. B.D. would like to thank Anjan Ananda Sen, Ranjini Mondol and Gaurav Goswami for helpful discussions.  
We acknowledge National Supercomputing Mission (NSM) for providing computing resources of `PARAM RUDRA' at P~G~Senapathy Center For Computer Resources, Play Field Ave, Indian Institute Of Technology, Chennai, Tamil Nadu 600036, which is implemented by C-DAC and supported by the Ministry of Electronics and Information Technology (MeitY) and Department of Science and Technology (DST), Government of India. 

\appendix

\section{Dark matter with Yukawa interactions}\label{app:UV_model}

Inspired by relativistic mean-field descriptions of nuclear matter at finite density, we consider a simple microscopic model for an asymmetric dark fermion $\chi$ interacting through a scalar mediator $\varphi$~\cite{Walecka:1974qa}. The scalar couples to the fermion bilinear $\bar\chi\chi$ with Yukawa coupling $g$. We also include a quartic scalar self-interaction, controlled by the coupling $\lambda$, which will play a role in regulating the attractive interaction at large density. The relevant interaction terms are
\begin{equation}
    \mathcal{L}_{\rm int}\supset-g\bar{\chi}\chi\varphi-\frac{\lambda}{4!}\varphi^4 .
    \label{eq:yukawa_lagrangian}
\end{equation}
Here we have chosen the conventional sign in which $\lambda>0$ corresponds to a stable repulsive scalar self-interaction in the potential. The scalar potential contains
\begin{equation}
    V(\varphi)=\frac{1}{2}m_\phi^2\varphi^2+\frac{\lambda}{4!}\varphi^4 .
\end{equation}

At finite density, the scalar density
\begin{equation}
    n_s \equiv \langle \bar{\chi}\chi\rangle~,
\end{equation}
is nonzero~\cite{Walecka:1974qa,Gresham:2018rqo,Garani:2022quc}. It sources a mean value of the scalar field, $\langle\varphi\rangle$, which in turn shifts the effective fermion mass. In the mean-field approximation, the scalar expectation value and the effective mass are determined by
\begin{eqnarray}
    g n_s &=& m_\phi^2 \langle\varphi\rangle + \frac{\lambda}{6}\langle\varphi\rangle^3 ,\label{eq:vev}\\[2mm]
    m_\chi^\ast&=&m_\chi- g\langle\varphi\rangle .
    \label{eq:effDMmass}
\end{eqnarray}
The scalar density for a zero-temperature degenerate fermion gas is
\begin{equation}
    n_s
    =
    \frac{m_\chi^\ast}{2\pi^2}
    \left[
    k_F\sqrt{k_F^2+m_\chi^{\ast 2}}
    -
    m_\chi^{\ast 2}
    \log
    \left(
    \frac{k_F+\sqrt{k_F^2+m_\chi^{\ast 2}}}{m_\chi^\ast}
    \right)
    \right] .
    \label{eq:scalarcondensate}
\end{equation}
Here $k_F$ is the Fermi momentum. For fixed microscopic inputs
\begin{equation}
    \{g,\lambda,m_\phi,m_\chi,k_F\},
\end{equation}
Eqs.~(\ref{eq:vev})--(\ref{eq:scalarcondensate}) are solved self-consistently for $\langle\varphi\rangle$ and $m_\chi^\ast$.

\subsection{Equation of state}

The zero-temperature equation of state is obtained from the sum of the degenerate-fermion contribution and the classical scalar-field contribution. In the mean-field approximation, the pressure and energy density are
\begin{eqnarray}
    P &=&-\frac{1}{2}m_\phi^2\langle\varphi\rangle^2-\frac{\lambda}{4!}\langle\varphi\rangle^4+\mathcal{F}_P(m_\chi,\langle\varphi\rangle,k_F),\label{eq:quartic_pressure}
    \\[2mm]
    \epsilon &=& \frac{1}{2}m_\phi^2\langle\varphi\rangle^2 + \frac{\lambda}{4!}\langle\varphi\rangle^4 + \mathcal{F}_E(m_\chi,\langle\varphi\rangle,k_F).
    \label{eq:quartic_energy}
\end{eqnarray}
The fermion functions are,
\begin{equation} \label{eq:degenerate_integrals}
\begin{aligned}
    \mathcal{F}_P(m_\chi,\langle\varphi\rangle,k_F) &=\frac{m_\chi^4}{3\pi^2} \int_0^{k_F/m_\chi}\frac{x^4\,dx}{\sqrt{x^2+\left(1-\frac{g\langle\varphi\rangle}{m_\chi}\right)^2}}~,
    \\[2mm]
    \mathcal{F}_E(m_\chi,\langle\varphi\rangle,k_F)&=\frac{m_\chi^4}{\pi^2} \int_0^{k_F/m_\chi} x^2 \sqrt{x^2+\left(1-\frac{g\langle\varphi\rangle}{m_\chi}\right)^2}\,dx~.
\end{aligned}
\end{equation}
Equivalently, these are the standard zero-temperature Fermi-gas integrals evaluated with the effective mass $m_\chi^\ast=m_\chi-g\langle\varphi\rangle$. The occupation function is
\begin{equation}
    f(p)=\Theta(\mu^\ast-E_p^\ast),
    \qquad
    \mu^\ast=\sqrt{k_F^2+m_\chi^{\ast 2}}~,
\end{equation}
with, $E_p^\ast=\sqrt{p^2+m_\chi^{\ast 2}}$. 
The resulting equation of state is obtained parametrically: for each value of $k_F$, one solves the mean-field equations for $\langle\varphi\rangle$ and $m_\chi^\ast$, computes $P$ and $\epsilon$, and then eliminate $k_F$ in favor of the mass density. This procedure gives the microscopic pressure-density relation against which the phenomenological equation of state can be compared.

\subsection{Mapping to phenomenological EoS}

We now clarify the relation between the microscopic Yukawa model and the phenomenological equation of state used in the main text, Eq.~\eqref{eq:eos_ph},
\begin{equation}
    P(\rho)=K\rho^{5/3}-\alpha\frac{(\rho/\rho_s)^2}{1+(\rho/\rho_s)^2}.
    \label{eq:eos_ph_app}
\end{equation}
This phenomenological form should not be viewed as an exact analytic reduction of the microscopic model. Rather, it is an effective parametrization designed to capture three qualitative features of an attractive finite-density interaction: a free degenerate Fermi-gas limit at low density, a negative pressure correction over an intermediate density range, and a recovery of the non-interacting scaling once the attractive contribution becomes regulated due to $\lambda$.

The cleanest analytic matching can be performed in the low-density, non-relativistic limit. For $ k_F \ll m_\chi$,
the scalar density reduces to the number density, $n_s \simeq n= k_F^3/(3\pi^2)$. At sufficiently small density, the scalar expectation value is also small, so that $m_\chi^\ast\simeq m_\chi$ and the $\lambda \langle \phi \rangle^3$ term in Eq.~(\ref{eq:vev}) is negligible. The gap equation then gives
\begin{equation}
    \langle\varphi\rangle\simeq\frac{g n}{m_\phi^2}.
    \label{eq:low_density_phi}
\end{equation}
Substituting this into the scalar part of the pressure gives
\begin{equation}
    \Delta P_{\rm micro} \simeq -\frac{1}{2}m_\phi^2\langle\varphi\rangle^2 \simeq -\frac{g^2}{2m_\phi^2}n^2 .
\end{equation}
Writing $n=\rho/m_\chi$, we get
\begin{equation}
    P_{\rm micro}\simeq K\rho^{5/3}-\frac{g^2}{2m_\phi^2m_\chi^2}\rho^2 .
    \label{eq:low_dens_corr}
\end{equation}
This expression has the same leading behavior as the low-density (non-relativistic) expansion of Eq.~(\ref{eq:eos_ph_app}),
\begin{equation}
    P_{\rm ph}(\rho) \simeq K\rho^{5/3}- \frac{\alpha}{\rho_s^2}\rho^2, \qquad \rho\ll \rho_s~.
\end{equation}
Thus, matching at low density gives
\begin{equation}
    \frac{\alpha}{\rho_s^2} \simeq\frac{g^2}{2m_\phi^2m_\chi^2}~.
    \label{eq:low_density_matching}
\end{equation}
This relation is the most direct connection between the phenomenological and microscopic descriptions. It shows that the combination $\alpha/\rho_s^2$ plays the role of an effective attractive two-body coupling in the non-relativistic, low density limit.

The interpretation of $\rho_s$ is more subtle. In the phenomenological equation of state, $\rho_s$ is the density scale at which the attractive correction stops growing as $\rho^2$ and begins to saturate. It is therefore not simply equal to $m_\phi$, $g$, or $\lambda$. Instead, $\rho_s$ should be understood as an effective density scale characterizing the onset of nonlinear finite-density effects that regulate the attractive force.

This distinction is important. If $\lambda=0$, the recovery of the equation of state from the attractive regime occurs only at relativistic densities. Indeed, for $ k_F \gg m_\chi^\ast$, the scalar density behaves as
$n_s\simeq m_\chi^\ast k_F^2/(2\pi^2)$~\cite{Gresham:2018rqo, Garani:2022quc}. For $\lambda=0$, the gap equation gives
\begin{equation}
    n_s = \frac{m_\phi^2}{g^2} (m_\chi-m_\chi^\ast).
\end{equation}
Combining the two relations implies that $m_\chi^\ast$ decreases with increasing $k_F$, approximately as $m_\chi^\ast\propto k_F^{-2}$ at asymptotically large density. The scalar density then saturates, and the scalar contribution to the pressure approaches a constant. Meanwhile the fermionic pressure grows as $k_F^4$, so the system eventually approaches the ultra-relativistic equation of state, $P\simeq \frac{\rho}{3}$.
Thus, in the absence of scalar self-interactions, the recovery from the pressure dip is tied to the onset of relativistic fermion dynamics.

The phenomenological EoS used in the main text is different in one important respect: its recovery occurs already in the non-relativistic regime. For $\rho\gg\rho_s$,
\begin{equation}
    -\alpha\frac{(\rho/\rho_s)^2}{1+(\rho/\rho_s)^2}\longrightarrow-\alpha ,
\end{equation}
so the derivative of the interaction piece vanishes. Consequently, the sound speed returns to the free Fermi-gas scaling,
\begin{equation}
    c_s^2=\frac{dP}{d\rho}\simeq\frac{5}{3}K\rho^{2/3}.
\end{equation}
This is why the phenomenological model can produce a localized suppression of the sound speed over a finite density interval, while reducing again to a non-interacting degenerate Fermi gas at higher density.

A positive scalar quartic provides one possible microscopic mechanism for such a non-relativistic recovery. In the regime where the cubic term in Eq.~\eqref{eq:vev} dominates over the quadratic term, one finds
\begin{equation}
    \frac{\lambda}{6}\langle\varphi\rangle^3\simeq g n\simeq\frac{g\rho}{m_\chi},
\end{equation}
and therefore
\begin{equation}
    \langle\varphi\rangle\simeq\left(\frac{6g\rho}{\lambda m_\chi} \right)^{1/3}.
\end{equation}
The scalar contribution to the pressure then scales as
\begin{equation}
    \Delta P_{\rm scalar}\sim-\lambda\langle\varphi\rangle^4 \propto -\rho^{4/3}.
\end{equation}
Since the non-relativistic degeneracy pressure scales as $P\propto \rho^{5/3}$, the fermion pressure eventually dominates again while the system is still non-relativistic. Thus, the quartic interaction can qualitatively reproduce the finite-density regulation that the phenomenological parameter $\rho_s$ is intended to capture.

However, the mapping is not exact. In the microscopic model with $\lambda>0$, the attractive contribution crosses over from a $\rho^2$ scaling to an approximate $\rho^{4/3}$ scaling. In the phenomenological EoS, by contrast, the attractive contribution saturates to a constant. Therefore, there is no unique closed-form identification of $\rho_s$ and $\alpha$ with $\lambda$, $g$, and $m_\phi$ beyond the low-density matching in Eq.~(\ref{eq:low_density_matching}). The phenomenological EoS should instead be understood as an effective description that preserves the relevant hydrostatic feature: a finite interval in density over which the sound speed is suppressed relative to the free Fermi gas.

This also clarifies which combination is most directly constrained by the dwarf-galaxy analysis. Differentiating Eq.~(\ref{eq:eos_ph_app}) gives the contribution to the sound speed,
\begin{equation}
    \Delta c_s^2(\rho)=-\frac{2\alpha}{\rho_s}\frac{\rho/\rho_s}{\left[1+(\rho/\rho_s)^2\right]^2}~.
    \label{eq:delta_cs_mapping}
\end{equation}
Thus $\alpha/\rho_s$ controls the maximum local suppression of the sound speed, while $\alpha/\rho_s^2$ controls the low-density two-body attractive pressure.

\subsection{Dark Matter Self-interaction constraints}

The finite-density EoS considered above is related to, but distinct from, conventional self-interacting EoS (e.g. ideal gas or non-interacting degenerate gas). Comparing hydrostatic solutions to data constrains how an attractive interaction modifies the bulk pressure and therefore the density profile of a degenerate DM halo. Standard SIDM bounds, on the other hand, constrain the rate of two-body momentum exchange in a dilute halo and are usually expressed in terms of a transfer or viscosity cross section per unit mass, $\sigma/m_\chi$.
A microscopic model such as Eq.~(\ref{eq:yukawa_lagrangian}) determines both the finite-density mean-field correction and the two-body scattering cross section, however, in general, the mapping between the two is density-dependent.

Observation of Bullet cluster constrain large self-scattering rates at typical velocities  $v\sim 1000~{\rm km\,s^{-1}}$. Upper bounds reads ~\cite{Spergel:1999mh,Harvey:2015hha,Bondarenko:2017rfu,Harvey:2018uwf,Sagunski:2020spe,DES:2023bzs}, 
\begin{equation}
\frac{\sigma_{\rm SI}}{m_\chi} \lesssim 0.5~{\rm cm^2\,g^{-1}}~.
    \label{eq:Bullet}
\end{equation}
At dwarf galaxy velocities, $v\sim 10\text{--}30~{\rm km\,s^{-1}}$,
larger cross sections are often phenomenologically allowed. In particular, the absence of clear evidence for gravothermal collapse in dwarf-sized halos gives an approximate upper bound of order
\begin{equation}
    \frac{\sigma_{\rm SI}}{m_\chi} \lesssim 100~{\rm cm^2\,g^{-1}},
\end{equation}
although this number is sensitive to halo age, concentration, baryonic effects, and the velocity dependence of the interaction~\cite{Tulin:2017ara,Adhikari:2022sbh, Fischer:2026ryr}.

For a Yukawa potential, the scattering cross section is generally velocity dependent. The relevant potential is schematically
\begin{equation}
    V(r)=-\frac{\alpha_\chi}{r}e^{-m_\phi r},\qquad \alpha_\chi\equiv \frac{g^2}{4\pi}.
\end{equation}
The scattering cross section depends on several dimensionless quantities, $
    \beta_Y = \frac{2\alpha_\chi m_\phi}{m_\chi v^2}$, $R_{\rm dB}=\frac{m_\chi v}{m_\phi}$, $R_B= \frac{\alpha_\chi m_\chi}{m_\phi}$.
Here $R_{\rm dB}$ compares the mediator range to the de Broglie wavelength, while $R_B$ compares the mediator range to the Bohr radius of the two-particle system. Depending on these quantities, scattering can occur in the Born, classical, or resonant regimes~\cite{Tulin:2013teo, Chu:2018faw, Chu:2018fzy, Chu:2019awd}. Consequently, a single velocity-independent number for $\sigma/m_\chi$ is generally insufficient.

We identify $\sigma_{\rm SI}$ with the transfer cross section when comparing with the bounds above, following appendix B1 of Ref.~\cite{Chu:2024gpe} and Ref.~\cite{Tulin:2013teo,Garani:2022quc}. In the Born regime, the cross section is perturbative and is given by
\begin{equation}
    \sigma_{\rm SI} =\frac{8\pi \alpha_\chi^2}{m_\chi^2 v^4}\left[\log\left(1+\frac{m_\chi^2 v^2}{m_\phi^2}\right)-\frac{m_\chi^2 v^2}{m_\phi^2+m_\chi^2 v^2}\right].
\end{equation}
This scaling shows why light mediators can be strongly constrained even for small couplings. Imposing bullet cluster constraint we get $\alpha_\chi<1.9\times 10^{-8} \left((100\,{\rm eV})/m_\chi\right)^{1/2}\left(m_\phi/(100\,{\rm eV})\right)^2$.

The connection with the phenomenological EoS is clearest in the low-density limit, where Eq.~(\ref{eq:low_density_matching}) gives
\begin{equation}
    \frac{\alpha}{\rho_s^2}
    \simeq
    \frac{g^2}{2m_\phi^2m_\chi^2}.
\end{equation}
Thus, a bound on the phenomenological combination $\alpha/\rho_s^2$ can be interpreted as a bound on the effective attractive mean-field coupling.
However, this is not identical to a bound on $\sigma_{\rm SI}/m_\chi$. Imposing $\alpha/\rho_s^2 <10^{-4}\,{\rm eV^{-4}}$, following results in presented in Sec.~\ref{sec:results}, we get $\alpha_\chi<1.6\times 10^{-7}\left(m_\chi/(100\,{\rm eV})\right)^2\left(m_\phi/(1\,{\rm meV})\right)^2$.

For this reason, any comparison between the MCMC constraints on the phenomenological EoS and conventional SIDM limits should be regarded as complementary. The dwarf galaxy LOS velocity dispersion analysis constrains the pressure support of a degenerate DM fluid. Cluster and halo-shape bounds constrain momentum exchange in dilute, collisionless systems. A complete microscopic analysis would require scanning the parameters $ \{g,m_\phi,\lambda,m_\chi\}$,
computing both the finite-density EoS and the velocity-dependent self-scattering cross section, and then imposing hydrostatic, kinematic, and SIDM constraints simultaneously. We leave this exploration for future work.

\subsection{Cosmological constraints}\label{app:cosmo}

The fermion masses preferred by the dwarf galaxy fits,
\begin{equation}
    m_\chi \sim \mathcal{O}(100)\,{\rm eV},
\end{equation}
are below the canonical lower bounds on thermal warm dark matter from Lyman-$\alpha$ forest measurements, which typically require
\begin{equation}
    m_{\rm WDM}\gtrsim {\rm few}\,{\rm keV}.
\end{equation}
However, these bounds assume a thermal relic with a Fermi--Dirac momentum distribution and negligible late-time self-interactions. Neither assumption is automatic in the class of models considered here.

First, if the dark matter is produced non-thermally, its momentum distribution can be colder than that of a thermal relic. The relevant free-streaming length scales approximately as
\begin{equation}
    \lambda_{\rm fs} \propto \frac{T_\chi}{m_\chi},
\end{equation}
where $T_\chi$ denotes the effective temperature of the dark sector~\cite{Garani:2021zrr}. A sub-keV particle with a sufficiently cold non-thermal distribution can therefore mimic the structure-formation behaviour of a heavier thermal warm-dark-matter candidate.

Second, Yukawa self-interactions can modify the early-time propagation of dark matter perturbations. If the dark-sector scattering rate
\begin{equation}
    \Gamma_\chi \sim n_\chi \sigma v
\end{equation}
exceeds the Hubble rate at early times, free-streaming is replaced by diffusion. The resulting cutoff in the matter power spectrum is then controlled not only by the particle mass, but also by the interaction rate and its velocity dependence. This is analogous in spirit to collisional damping in baryon--photon or dark-sector fluids.

Consequently, the mapping between $m_\chi$ and the suppression of small-scale structure is model-dependent. The standard Lyman-$\alpha$ bounds on thermal warm dark matter do not directly apply to a non-thermal, self-interacting fermionic dark sector~\cite{Carena:2021bqm,Egana-Ugrinovic:2021gnu,Garani:2022yzj}. A dedicated cosmological analysis would require specifying the production mechanism, the dark-sector temperature, and the velocity-dependent self-interaction rate, and then computing the resulting linear matter power spectrum.

\section{Maxwell's Construction}\label{app:maxwell_construction}
\begin{figure*}[t!]
    \centering
          \includegraphics[width=0.47\textwidth]{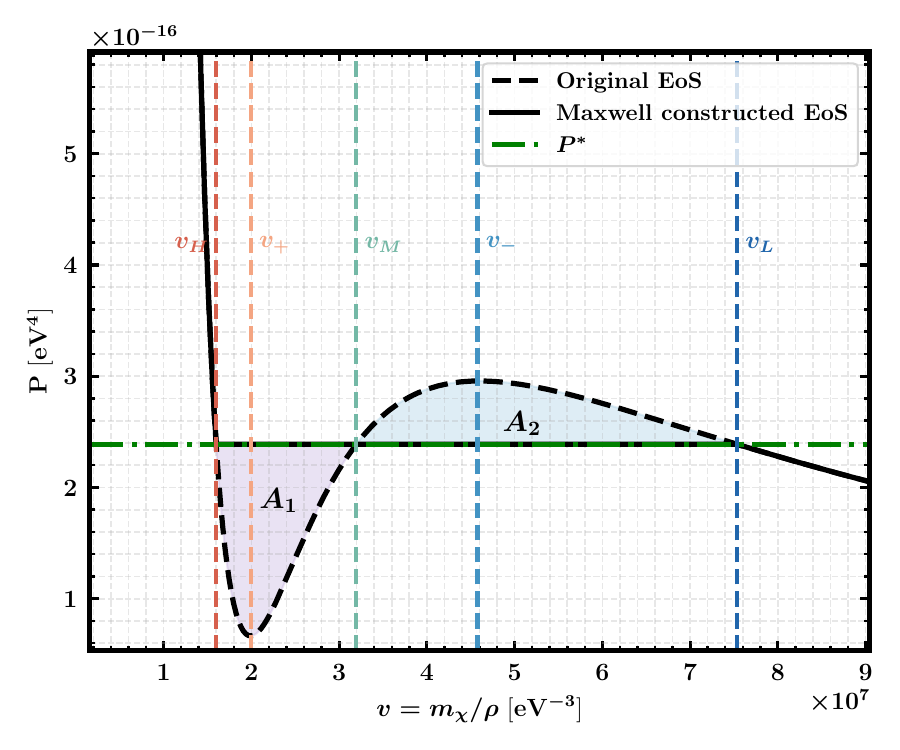}
        \includegraphics[width=0.47\textwidth]{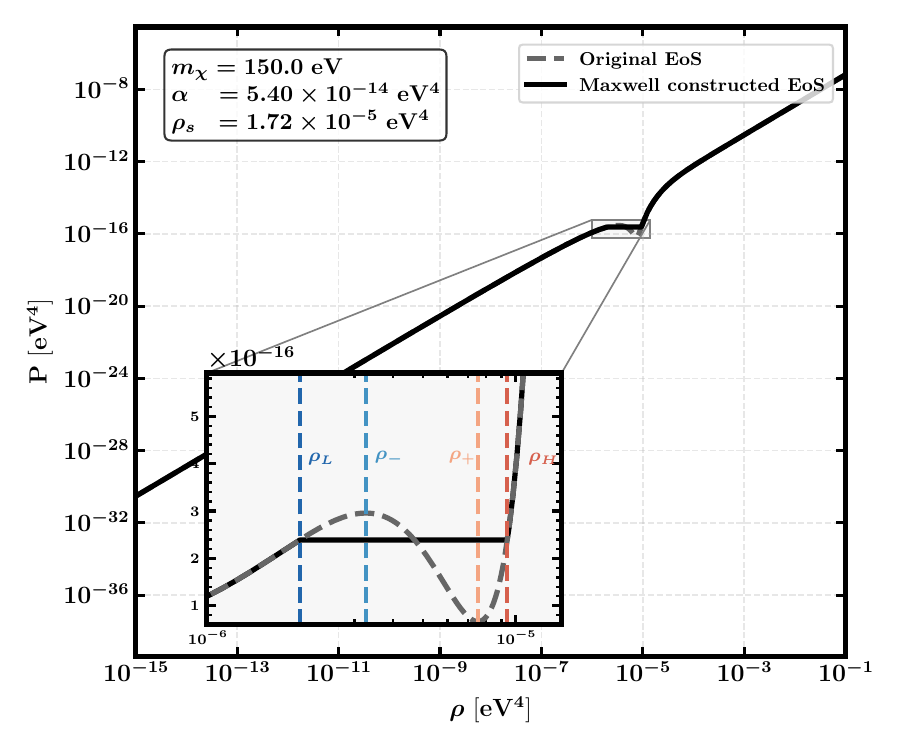}
    \caption{{\it Left:} An example of a Maxwell-construction in the $P$-$v$ plane, using the equal area rule. {\it Right:} Resultant Maxwell constructed EoS in the $P-\rho$ plane. }
    \label{fig:eos_maxwell}
\end{figure*}

 When the parameters $K$, $\alpha$ and $\rho_s$ are such that the EoS develops a non-monotonic region, mechanical stability requires that the unphysical segment be replaced by a constant-pressure coexistence plateau, determined by Maxwell's construction. Below we will summarize Maxwell's construction in the form applicable to EoS described by Eq.~\eqref{eq:eos_ph}.
 
The boundaries of the mechanically unstable region are defined by the spinodal
condition $dP/d\rho = 0$:
\begin{equation}
    \frac{dP}{d\rho} = \frac{5}{3} K \rho^{2/3} - \frac{2\alpha\rho_s^2 \rho}{\left(\rho_s^2+\rho^{2}\right)^2}  = 0.
    \label{eq:spinodal}
\end{equation}
This equation admits two roots $\rho_{-}$ and $\rho_{+}$ (with $\rho_<\rho_{+}$), between which $dP/d\rho < 0$ and the system is mechanically unstable. At zero temperature the Gibbs-Duhem relation reads
\begin{equation}
    dP = \rho\, d\mu,
\end{equation}
so the chemical potential can be reconstructed from the EOS via
\begin{equation}
    \mu(\rho) = \frac{P(\rho)}{\rho} + \frac{\mathcal{E}(\rho)}{\rho},
\end{equation}
where the energy density $\mathcal{E}(\rho)$ satisfies
\begin{equation}
    P = \rho^{2}\frac{d\!\left(\mathcal{E}/\rho\right)}{d\rho}
    \implies
    \mathcal{E}(\rho) = \rho\int \frac{P(\rho)}{\rho^{2}}\,d\rho.
\end{equation}

Phase coexistence between a low-density phase at $\rho_{L}$ and a high-density phase at $\rho_{H}$ (with $\rho_{L}<\rho_{-}<\rho_{+}<\rho_{H}$) requires simultaneous mechanical and chemical equilibrium:
\begin{align}
    P(\rho_{L}) &= P(\rho_{H}) = P^{*}, \label{eq:mech}\\
    \mu(\rho_{L}) &= \mu(\rho_{H}). \label{eq:chem}
\end{align}
 
Combining equations~\eqref{eq:mech} and \eqref{eq:chem} yields the equal-area rule. From $\mu(\rho_{H})-\mu(\rho_{L})=0$ one has
\begin{equation}
    \int_{\rho_{L}}^{\rho_{H}} \frac{1}{\rho}\frac{dP}{d\rho}\,d\rho = 0.
\end{equation}
Integrating by parts and using $v\equiv m_\chi/\rho$ as the specific volume gives
\begin{equation}
    P^{*}\!\left(v_{L}-v_{H}\right) = \int_{v_{H}}^{v_{L}} P(v)\,dv,
    \label{eq:equalarea}
\end{equation}
which states that the horizontal tie-line at $P=P^{*}$ must divide the $P$--$v$ loop into two regions of equal area, i.e. the well-known equal area rule: 
\begin{equation}
    \int_{v_L}^{v_M} \left[P^* - P(v)\right]dv = \int_{v_M}^{v_H} \left[P(v) - P^*\right]dv\ , 
\end{equation}
where $v_M$ is the middle root of $P(v) = P^*$ in the unstable region where $\mathrm{d}P/\mathrm{d}v > 0$. 
Note that $v_L = m_\chi/\rho_H$ and $v_H = m_\chi/\rho_L$. 
An example of using the equal area rule is shown in the left panel of figure~\ref{fig:eos_maxwell}.
Equivalently, in the
$P$--$\rho$ plane,
\begin{equation}
    \int_{\rho_{L}}^{\rho_{M}}\!\bigl[P(\rho)-P^*\bigr]\frac{d\rho}{\rho^{2}}
    =
    \int_{\rho_{M}}^{\rho_{H}}\!\bigl[P^*-P(\rho)\bigr]\frac{d\rho}{\rho^{2}},
\end{equation}
where $\rho_M$ is again the root of $P(\rho) = P^*$, in the unstable region, that separates the two lobes. 
For the EoS given in Eq.~\eqref{eq:eos_ph} the required integral evaluates
analytically:
\begin{equation}
    \int \frac{P(\rho)}{\rho^{2}}\,d\rho
    = \frac{3}{2}K\,\rho^{2/3} -\frac{\alpha}{\rho_s} \arctan\left(\frac{\rho}{\rho_s}\right) + C.
\end{equation}
The equal-area condition~\eqref{eq:equalarea} then reads explicitly
\begin{eqnarray}
    P^{*}\left(\frac{1}{\rho_{H}}-\frac{1}{\rho_{L}}\right)
    &&= \frac{3}{2}K\left(\rho_{H}^{2/3}-\rho_{L}^{2/3}\right)\nonumber \\
    & &- \frac{\alpha}{\rho_s}\!\left[\arctan\left(\frac{\rho_{H}}{\rho_s}\right)-\arctan\left(\frac{\rho_{L}}{\rho_s}\right)\right]\nonumber~. \\
    \label{eq:Ps_expl}
\end{eqnarray}

The three unknowns $\{\rho_{L},\,\rho_{H},\,P^{*}\}$ are determined by Eqs.~\eqref{eq:mech} and~\eqref{eq:Ps_expl}:
\begin{align}
    K \rho_{L}^{5/3} - \alpha\frac{\rho_{L}^{2}}{\rho_s^2+\rho_{L}^{2}} &= P^{*},
    \label{eq:sys1}\\
    K \rho_{H}^{5/3} - \alpha \frac{\rho_{H}^{2}}{\rho_s^2+\rho_{H}^{2}} &= P^{*},
    \label{eq:sys2}\\
    P^{*}\!\left(\frac{1}{\rho_{H}}-\frac{1}{\rho_{L}}\right)
    &= \frac{3}{2}K\!\left(\rho_{H}^{2/3}-\rho_{L}^{2/3}\right)\nonumber \\
    &-\frac{\alpha}{\rho_s}\!\left[\arctan\left(\frac{\rho_{H}}{\rho_s}\right) -\arctan\left(\frac{\rho_{L}}{\rho_s}\right)\right].
    \label{eq:sys3}
\end{align}
This nonlinear system can be solved numerically, with the densities $\rho_{\pm}$ serving as initial brackets for $\rho_{L}$ and $\rho_{H}$ respectively. In practice, we solve the equivalent system in the $P-v$ plane, where $v = m_\chi/\rho$.
  
The physical EoS, with the unstable piece replaced by the coexistence plateau, is
\begin{equation}
    P_{\mathrm{MC}}(\rho) =
    \begin{cases}
        \displaystyle K\rho^{5/3} - \alpha \frac{\rho^{2}}{\rho_s^2+\rho^{2}} & \rho < \rho_{L},\\[8pt]
        P^{*} & \rho_{L} \leq \rho \leq \rho_{H},\\[6pt]
        \displaystyle K\rho^{5/3} - \alpha\frac{\rho^{2}}{\rho_s^2+\rho^{2}} & \rho > \rho_{H}.
    \end{cases}
    \label{eq:MC}
\end{equation}

\begin{figure}[htb!]
    \centering
    \includegraphics[width=\linewidth]{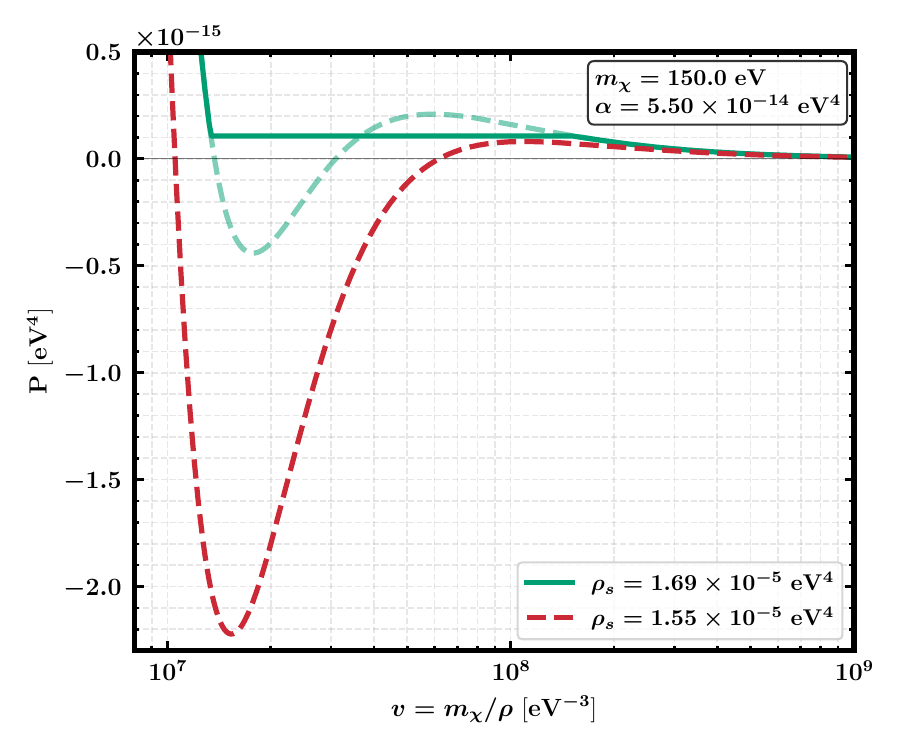}
    \caption{\justifying In this plot we fix $m_\chi = 150\ \text{eV}$ and $\alpha = 5.5\times 10^{-14}\ \text{eV}^4$. 
    The green curve (dashed) shows pressure as a function of the specific volume for $\rho_s = 1.69\times 10^{-5}\ \text{eV}^4$ where the pressure $P(v_+) < 0$, while the green (solid) shows the Maxwell-constructed curve. The dashed red curve exhibits a dip sufficiently deep that no $P^* > 0$ line yields equal lobe areas, and the Maxwell construction fails.}
    \label{fig:maxwell_cases}
\end{figure}

This represents a phase transition in the dark matter fluid, with $P^{*}$ as the transition pressure and $\rho_{L}$, $\rho_{H}$ as the coexisting phase densities. Physically, interpretable as the boundary between a diffuse dark matter halo phase and a dense dark matter core phase within dwarf galaxies.

As an illustrative example we consider the following parameters, $m_\chi = 150\ \mathrm{eV}$, $\alpha = 5.4\times 10^{-13}\ \mathrm{eV}^4$, $\rho_s = 1.72\times 10^{-5}\ \mathrm{eV}^4$, the Maxwell-constructed results for which are shown in figure~\ref{fig:eos_maxwell}.

However, it also worth noting the following two cases: 
(a) For certain parameters the pressure can be negative, and yet Maxwell construction can be done for a $P^* > 0$ since the dip is shallow enough in the P-V plane to satisfy the equal-area rule.
(b) On the other hand, for certain parameters for a small enough $\rho_s$ or large enough $\alpha$, the dip becomes deep enough that no $P^* > 0$ can satisfy the equal-area rule in the P-V plane. 
Here, no Maxwell construction is possible (see also Appendix~F of Ref.~\cite{Garani:2022quc}).
We illustrate both of these examples in figure~\ref{fig:maxwell_cases}.~\footnote{The $x$-axis is plotted on a log scale for clarity, which distorts the apparent lobe areas; the equal-area construction is exact in linear $v$.}

\section{MCMC results}\label{app:mcmc_results}

\begin{figure*}[htb!]
    \centering
    \includegraphics[width=0.47\linewidth]{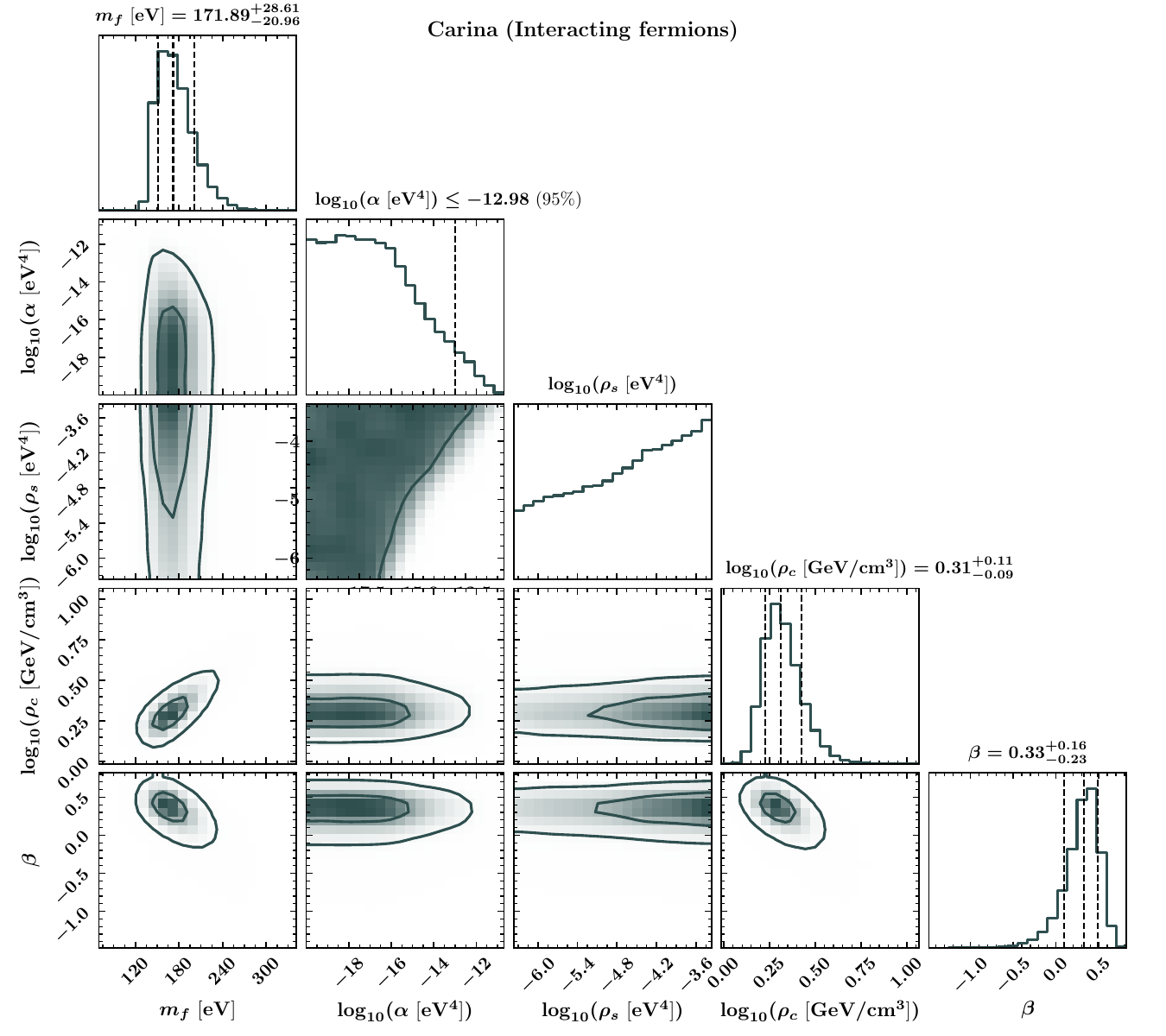}
    \includegraphics[width=0.47\linewidth]{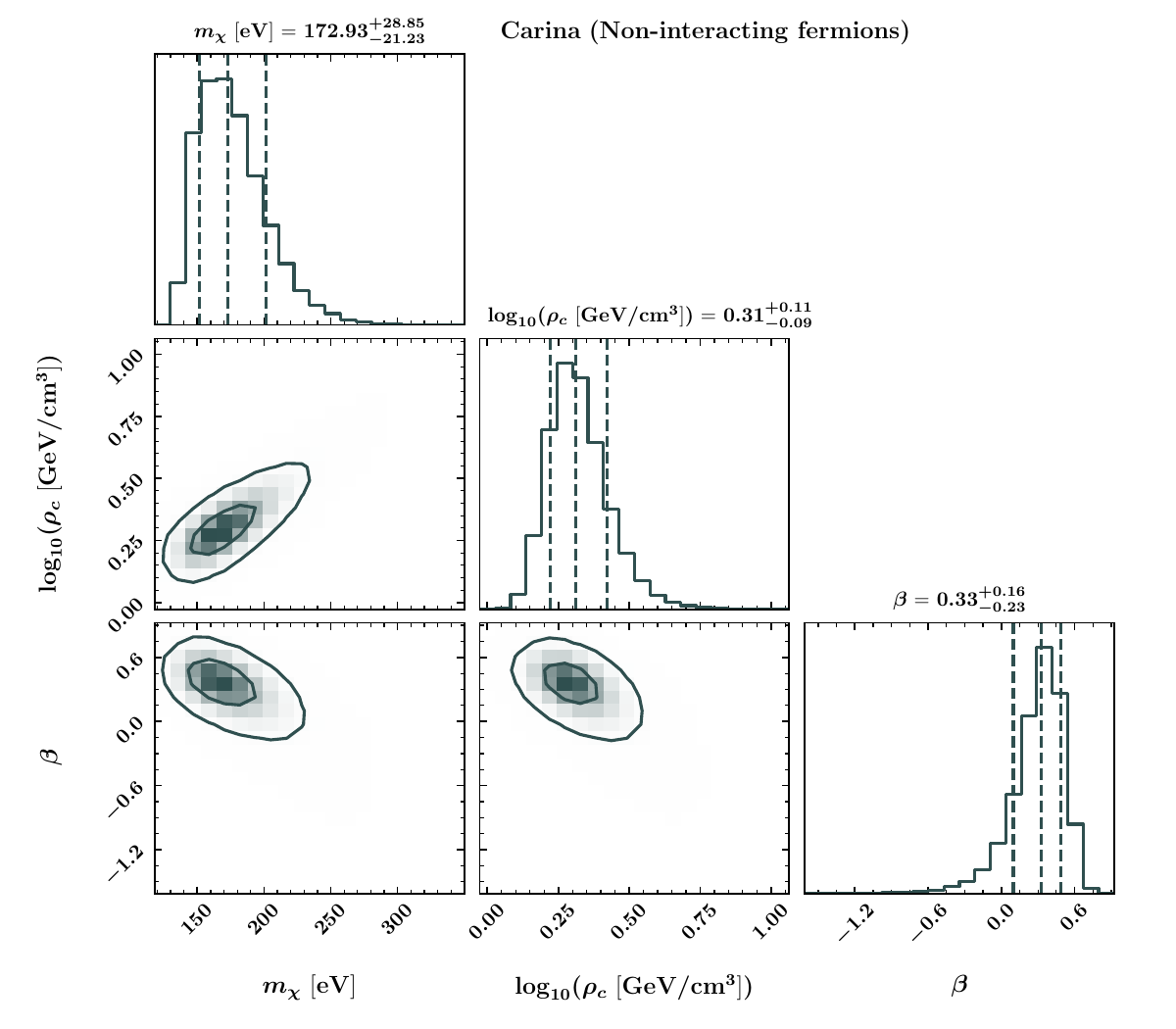}\\
    \includegraphics[width=0.47\linewidth]{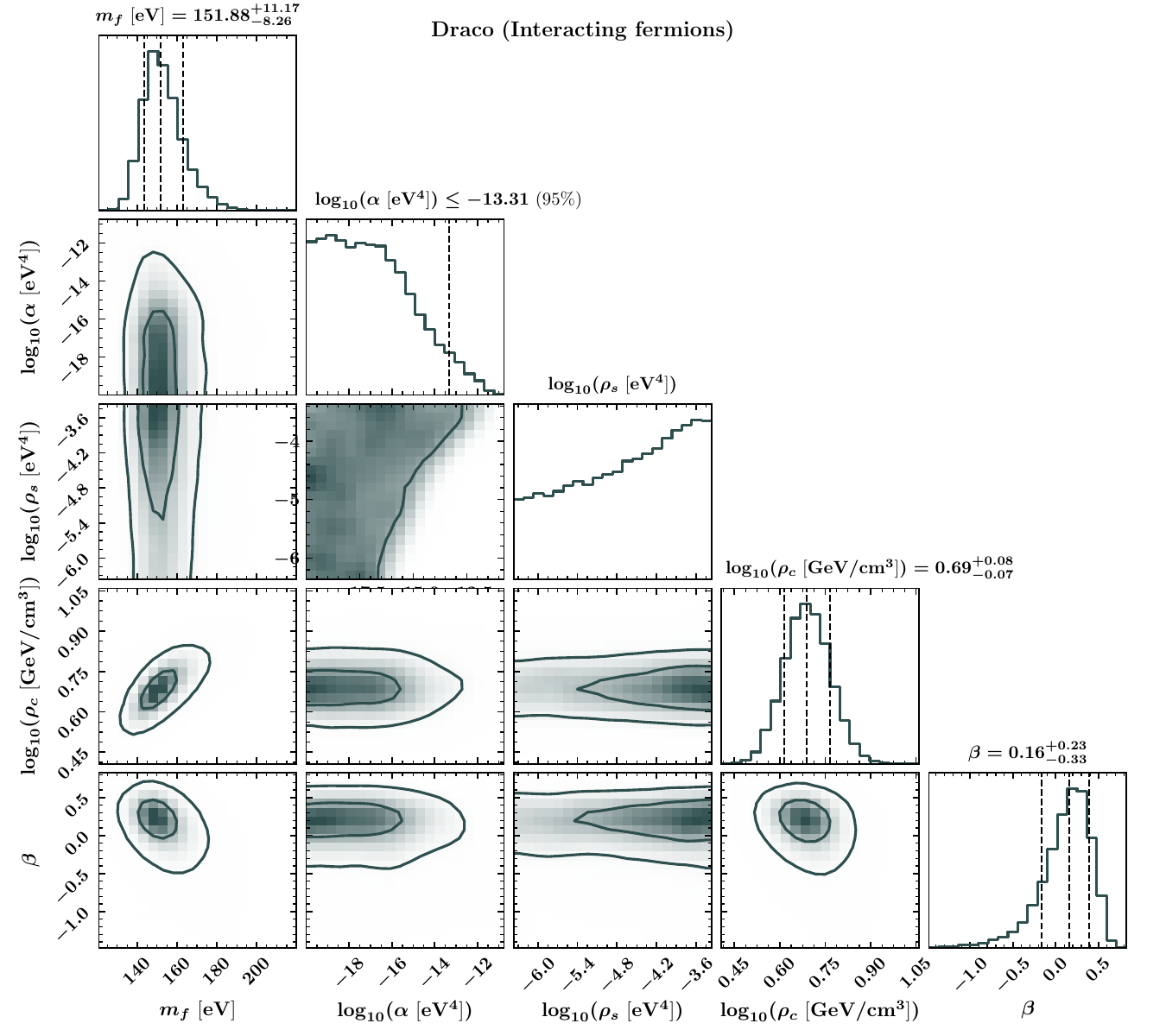}
    \includegraphics[width=0.47\linewidth]{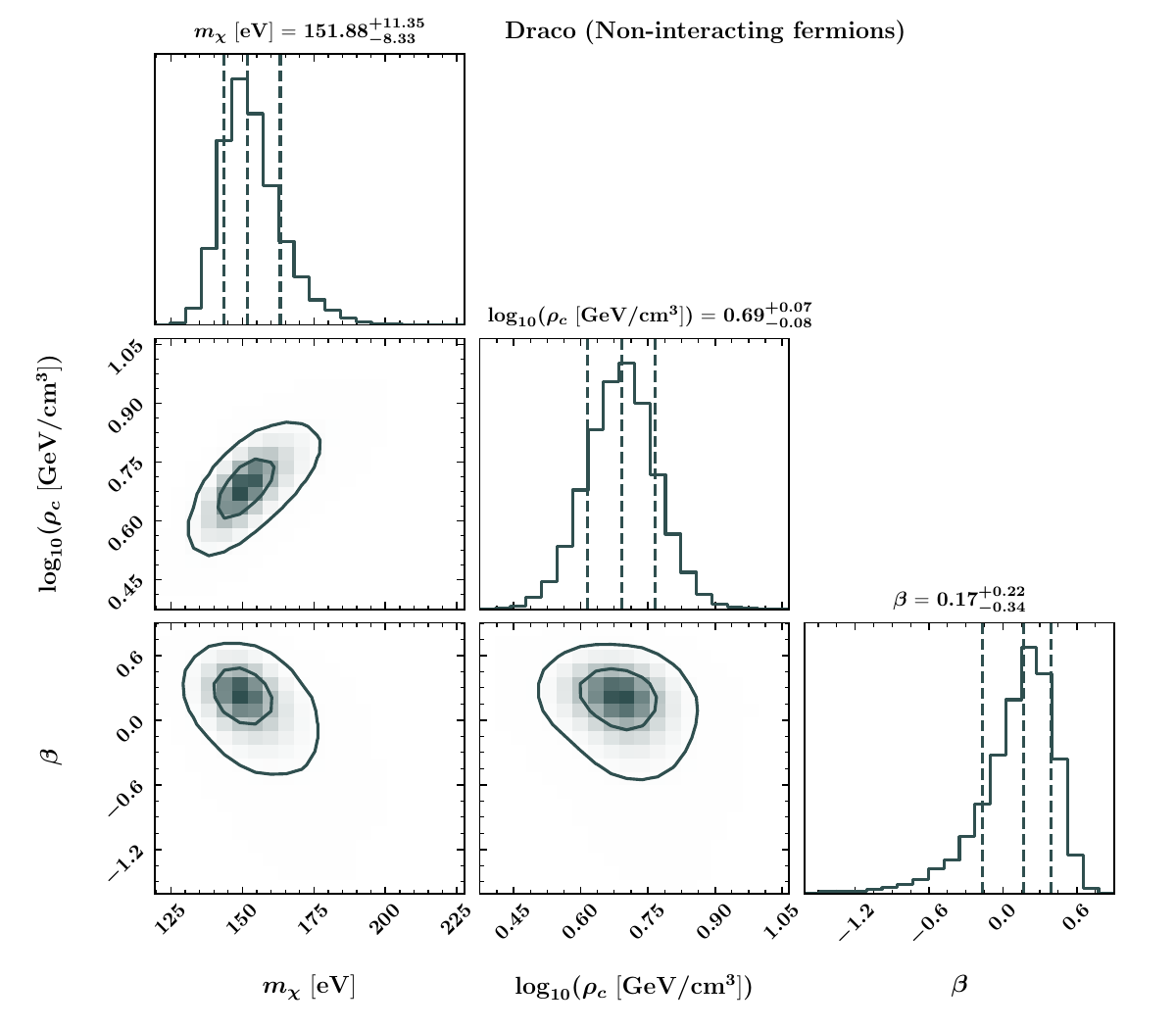}
    \caption{The same as figure~\ref{fig:fornax_leoII_post} for Carina and Draco.}
    \label{fig:carina_draco_post}
\end{figure*}

\begin{figure*}[h]
    \centering
    \includegraphics[width=0.47\linewidth]{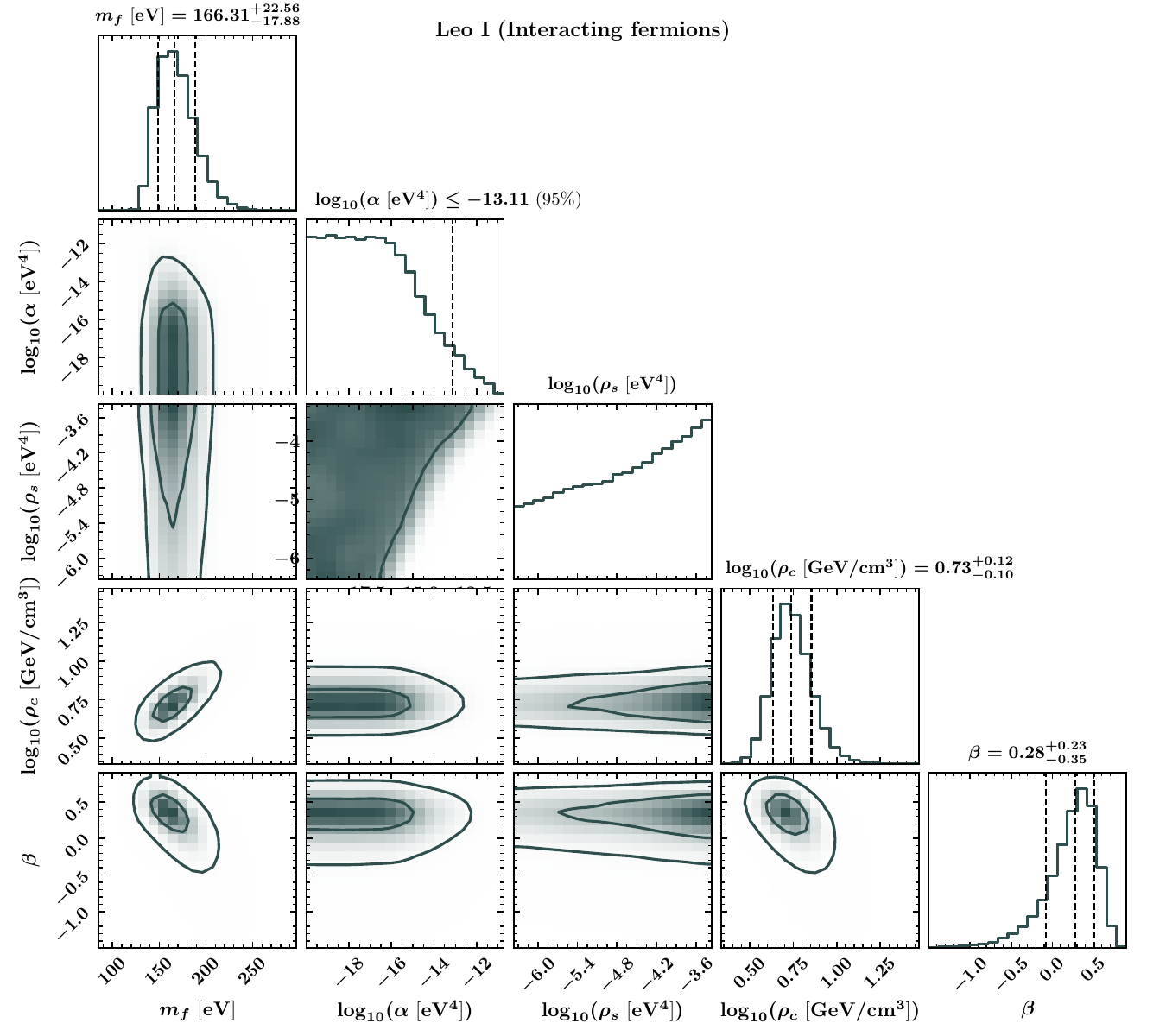}
    \includegraphics[width=0.47\linewidth]{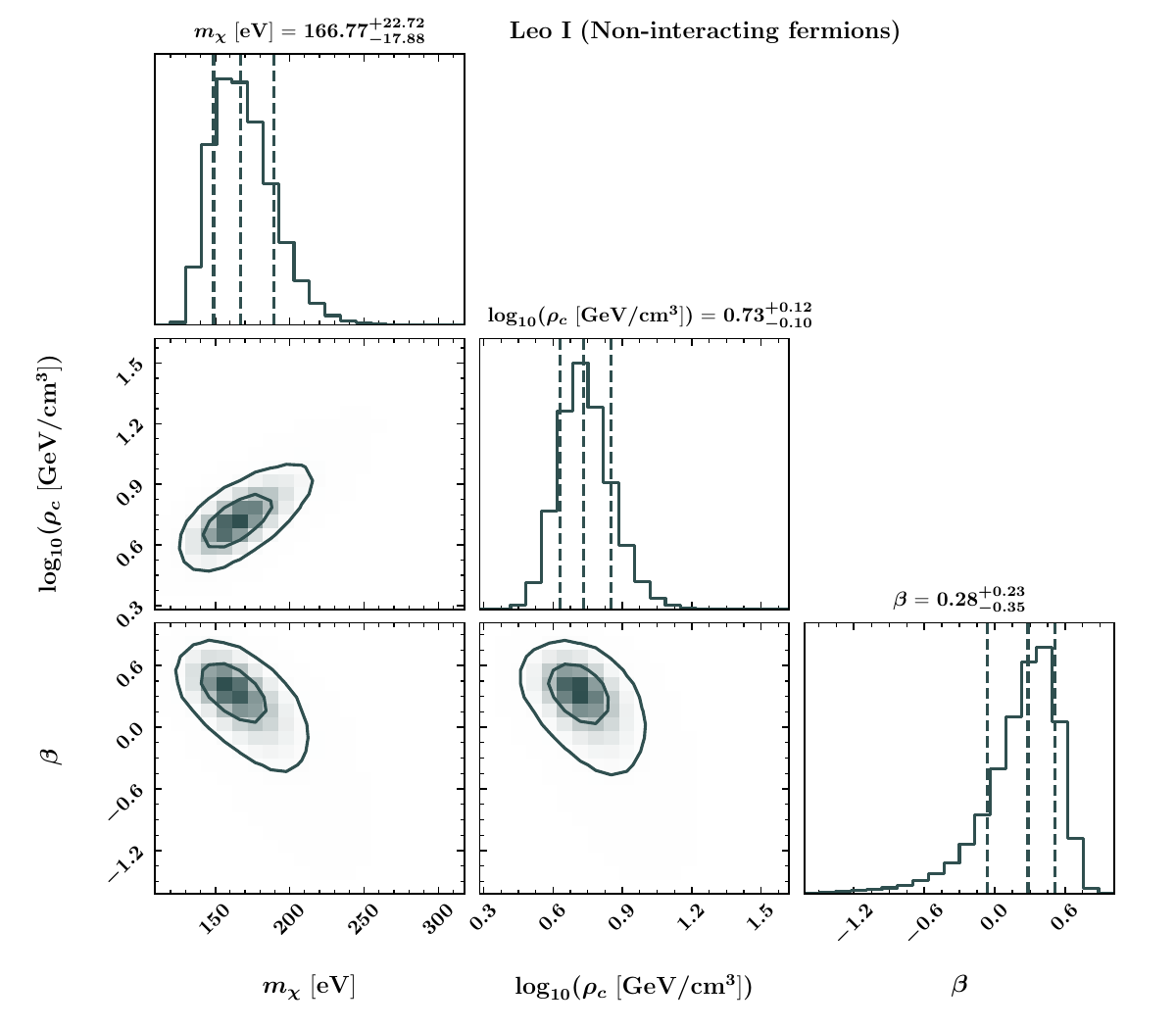}\\
    \includegraphics[width=0.47\linewidth]{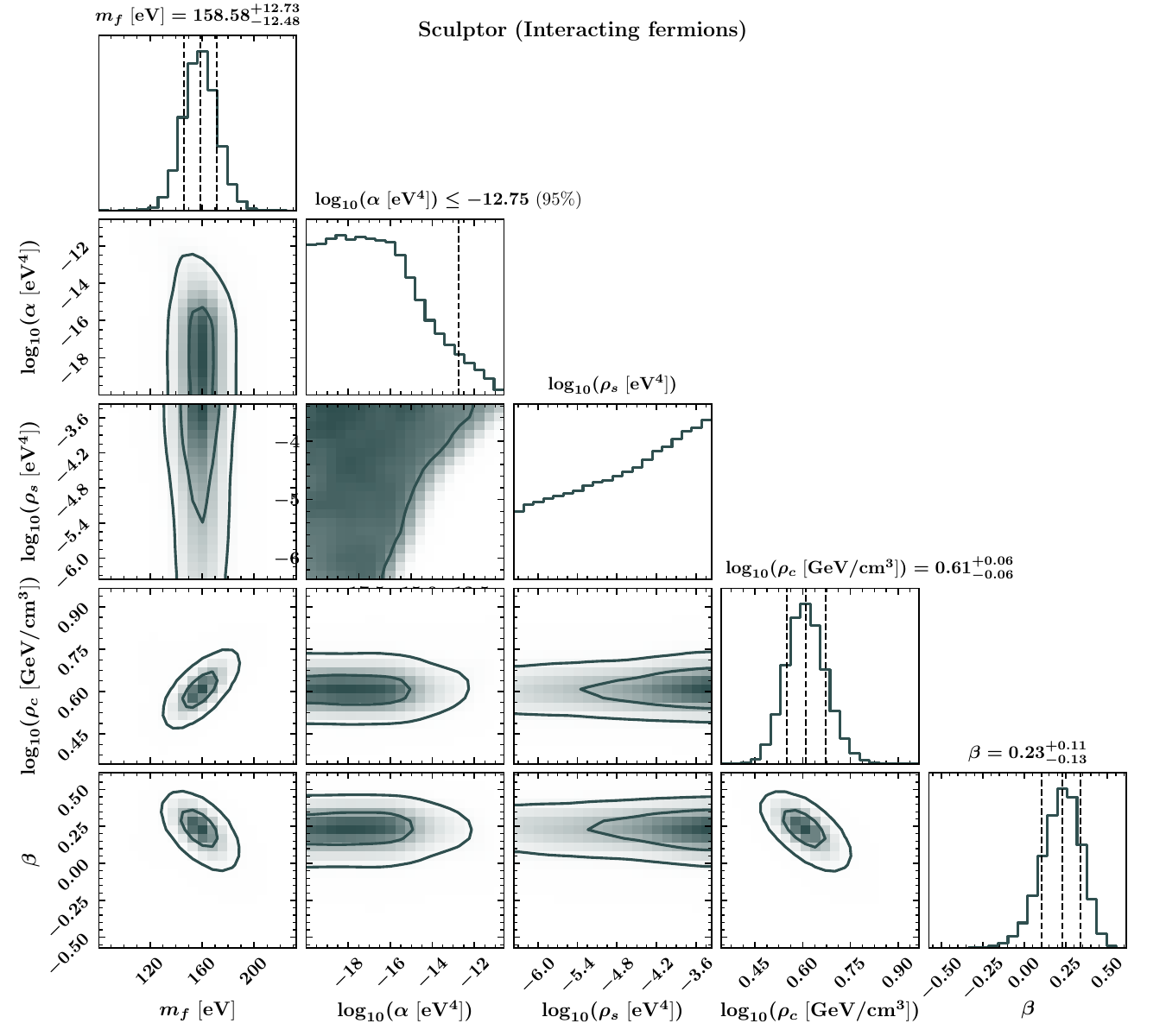}
    \includegraphics[width=0.47\linewidth]{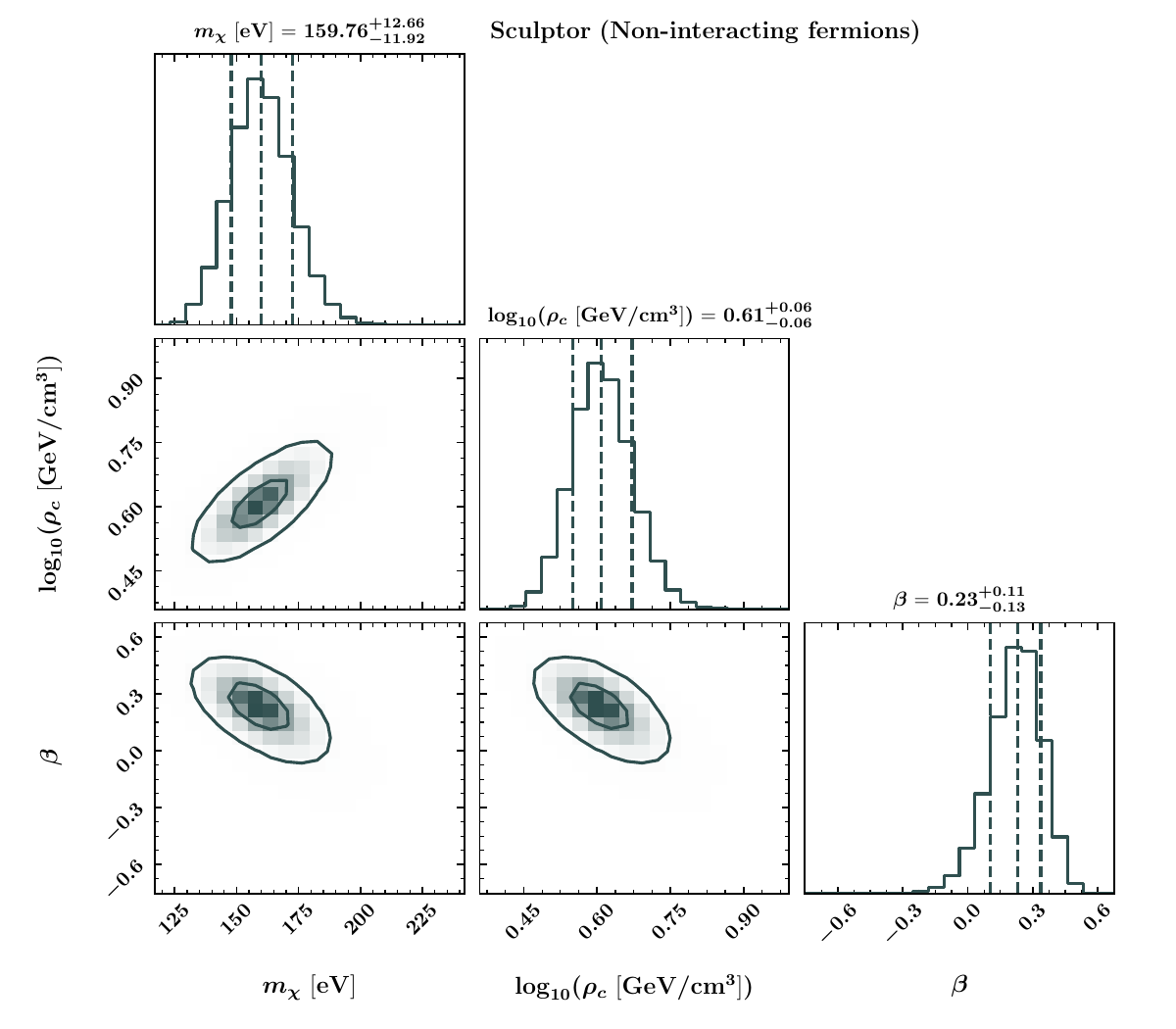}
    \caption{The same as figure~\ref{fig:fornax_leoII_post} for Leo I and Sculptor.}
    \label{fig:leoI_sculptor_post}
\end{figure*}

\begin{figure*}[h]
    \centering
    \includegraphics[width=0.47\linewidth]{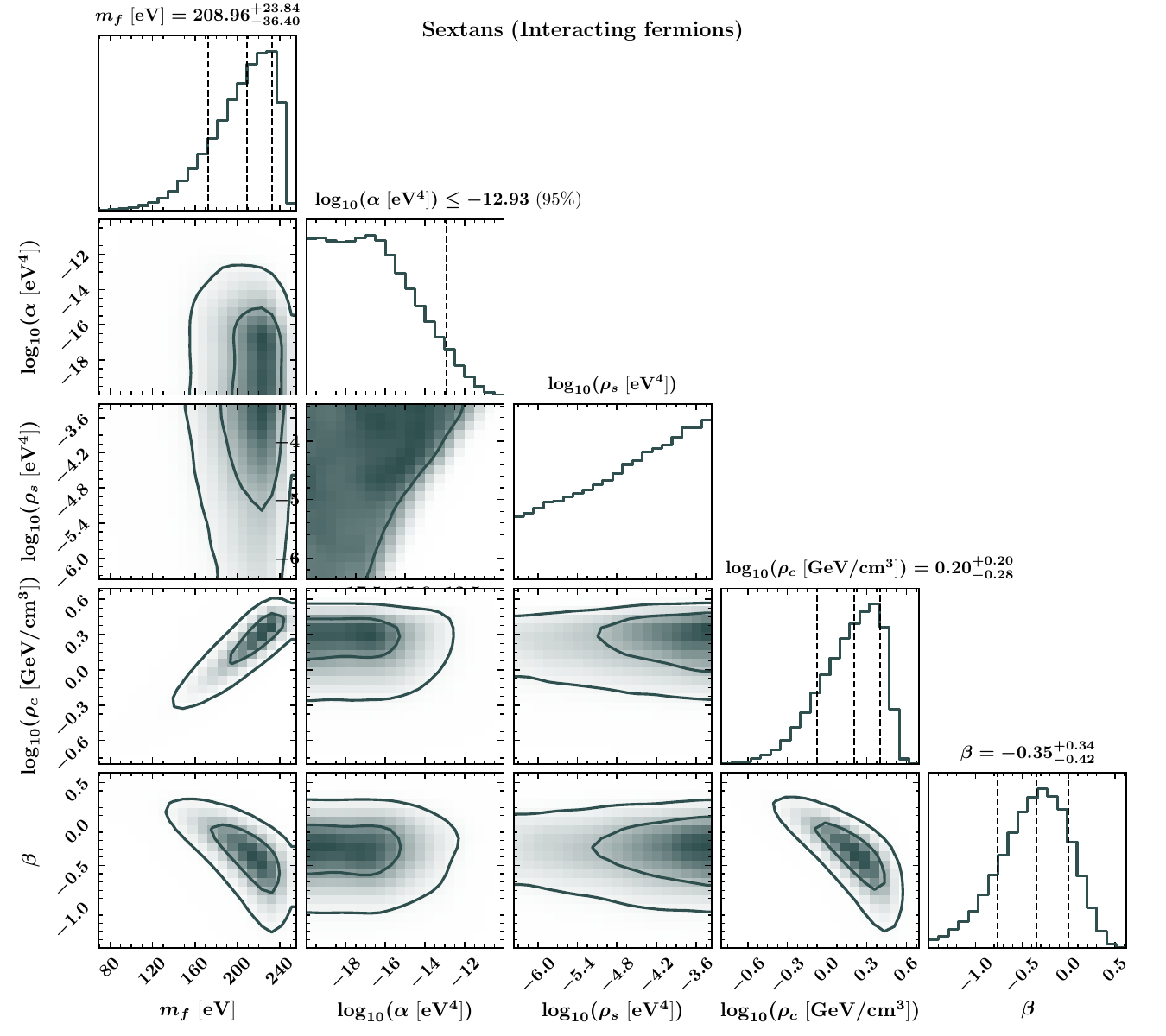}
    \includegraphics[width=0.47\linewidth]{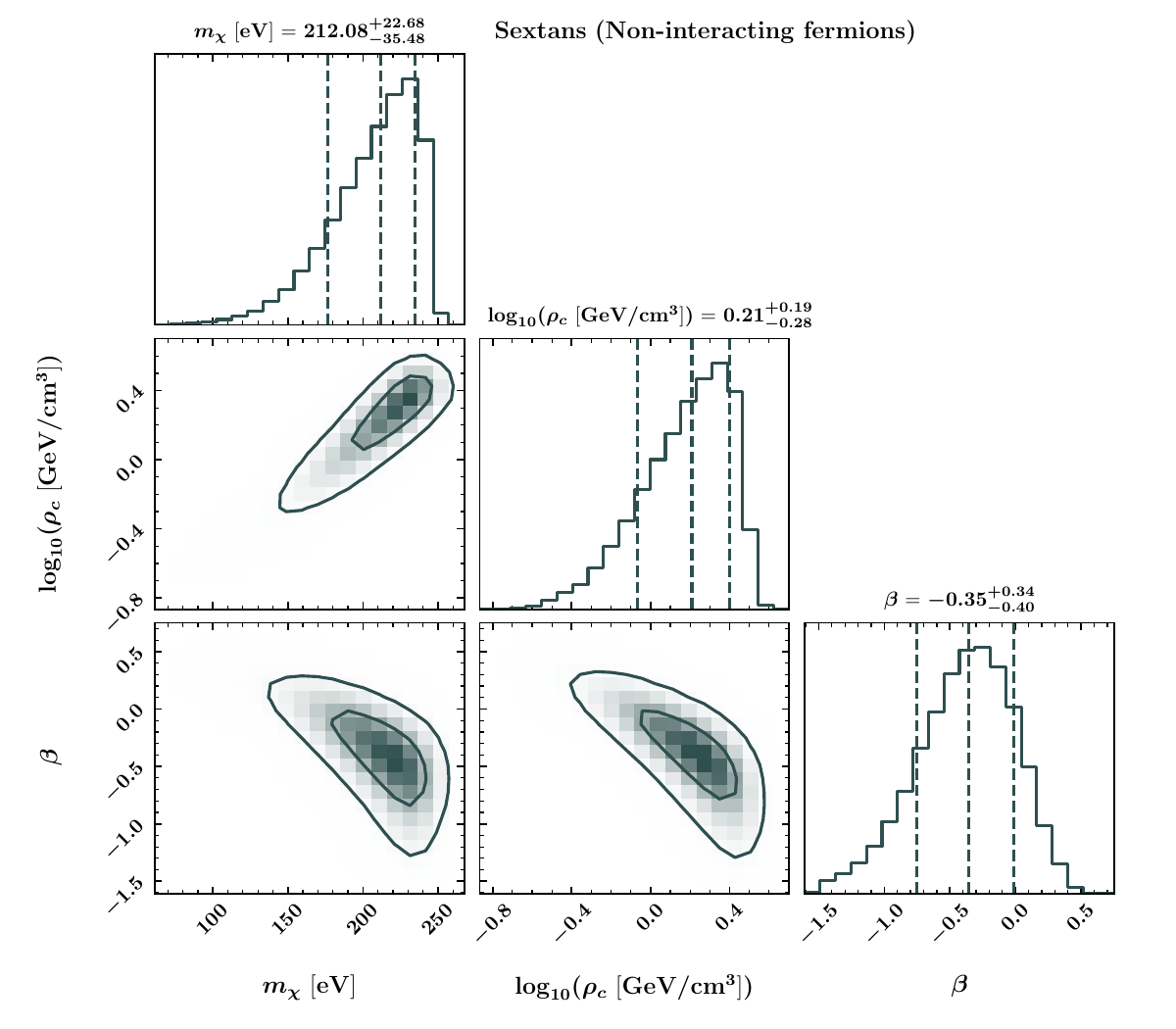}\\
    \includegraphics[width=0.47\linewidth]{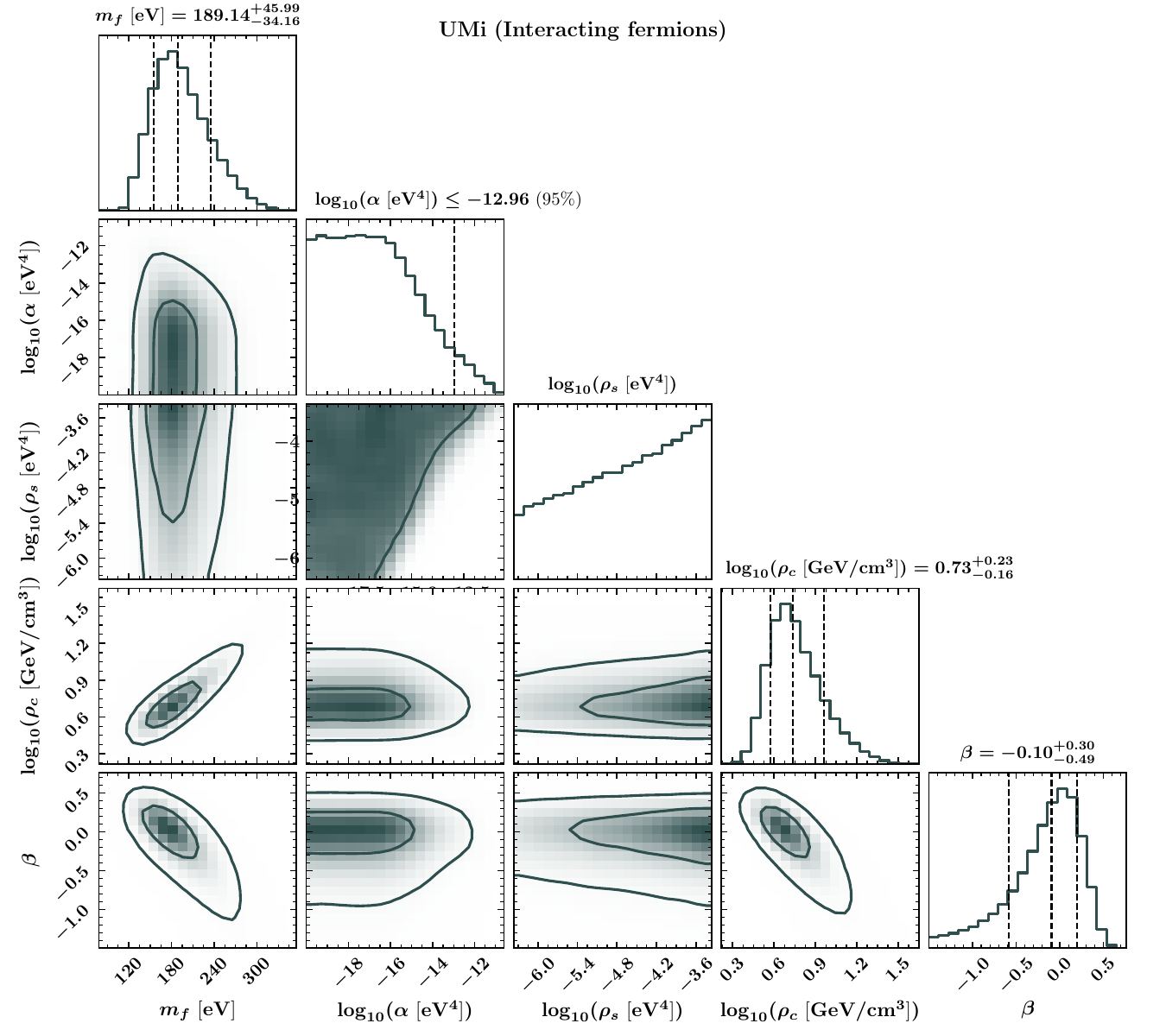}
    \includegraphics[width=0.47\linewidth]{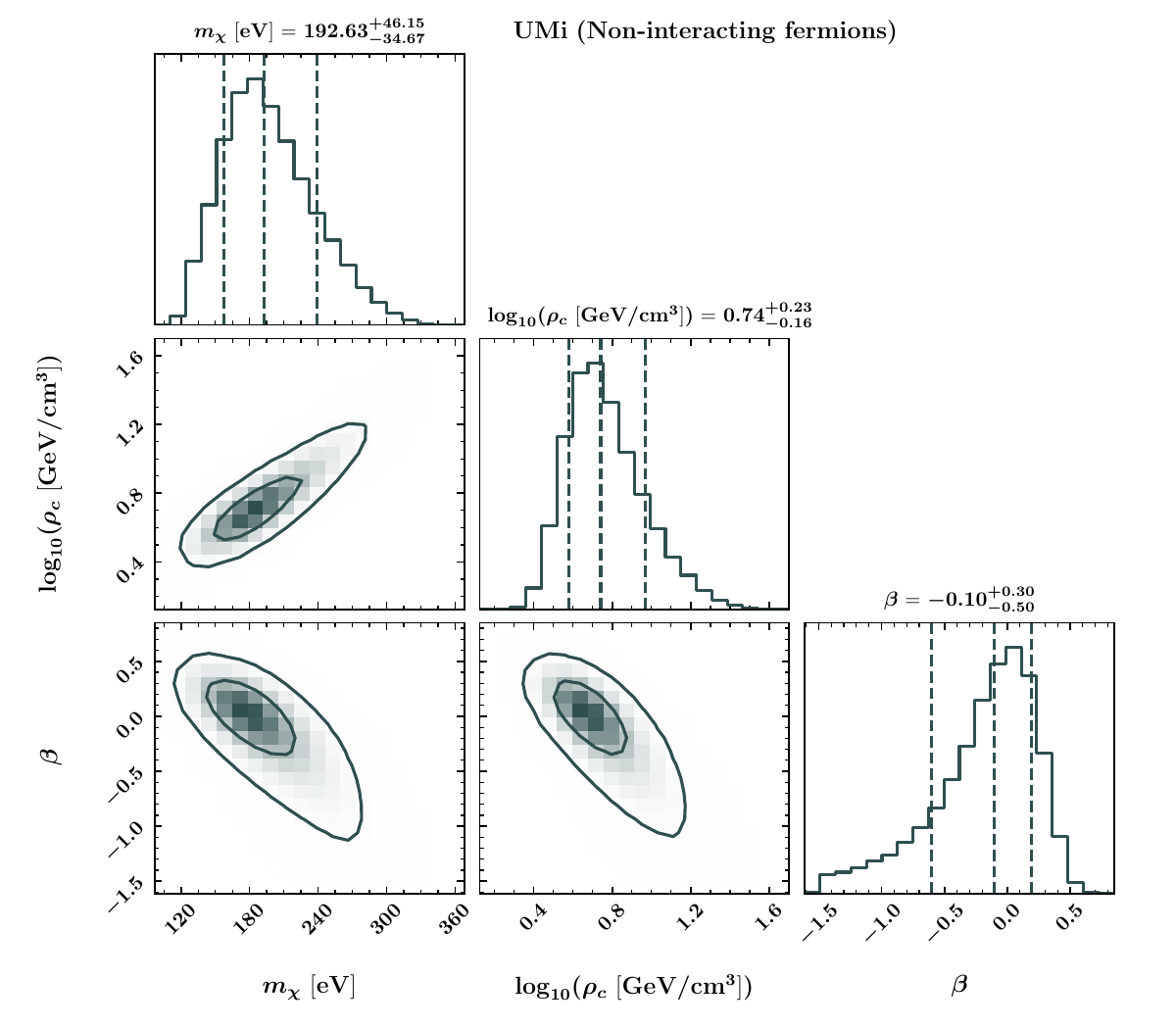}
    \caption{The same as figure~\ref{fig:fornax_leoII_post} for Sextans and Ursa Minor.}
    \label{fig:sextans_umi_post}
\end{figure*}

This appendix collects the full set of posterior distributions obtained from the MCMC runs. In the main text we have shown representative examples for Fornax and Leo~II, which illustrate the typical constraints on interacting and non-interacting EoS. Here, we provide the corresponding corner plots for the remaining six classical dwarf spheroidals: Carina, Draco in Fig.~\ref{fig:carina_draco_post}, Leo~I, Sculptor in Fig.~\ref{fig:leoI_sculptor_post}, and Sextans, Ursa Minor in Fig.~\ref{fig:sextans_umi_post}. In each case, the left panel shows the five parameters for interacting phenomenological EoS, with parameters $(m_\chi,\log_{10}\alpha,\log_{10}\rho_s,\log_{10}\rho_c,\beta)$, while the right panel shows the three parameter non-interacting degenerate Fermi-gas model, with parameters $(m_\chi,\log_{10}\rho_c,\beta)$. These plots demonstrate that the qualitative conclusions discussed in Sec.~\ref{sec:results} are common across the sample: the fermion mass, central density, and anisotropy posteriors are broadly consistent between the two models, while the interaction parameters are not localized around a preferred non-zero value. Instead, the data mainly impose upper limits on the strength of the attractive interaction, with the allowed region extending toward the effectively non-interacting regime.

\bibliographystyle{apsrev4-1}
\bibliography{biblio}

\end{document}